\documentclass[sigconf, screen]{acmart}
\usepackage{booktabs}
\usepackage{multirow}
\usepackage{tabularx}
\usepackage{threeparttable}
\usepackage{array}
\usepackage{colortbl}
\usepackage{microtype}
\definecolor{lavHead}{HTML}{B6B3D6}
\definecolor{lavA}{HTML}{CFCCE3}
\definecolor{lavB}{HTML}{D5D3DE}
\definecolor{neuGray}{HTML}{D5D1D1}
\definecolor{corLight}{HTML}{F6DFD6}
\definecolor{corMid}{HTML}{F8B2A2}
\definecolor{corAccent}{HTML}{F1837A}
\definecolor{corDark}{HTML}{E9687A}
\newcommand{\lavrow}{\rowcolor{lavHead}}          
\newcommand{\correow}{\rowcolor{lavHead}}         
\newcommand{\neurow}{\rowcolor{neuGray}}          

\AtBeginDocument{%
  }

\setcopyright{none}
\copyrightyear{2027}
\acmYear{2027}
\acmDOI{}
\acmConference[KDD '27]{The 33rd ACM SIGKDD Conference on Knowledge Discovery and Data Mining}{August 1--5, 2027}{San Jose, CA, USA}
\acmISBN{}

\def\sysname{\textsc{BATON}}

\title[\sysname]{\sysname: A Multimodal Benchmark for Bidirectional Automation Transition Observation in Naturalistic Driving}

\author{Yuhang Wang}
\email{yuhangw@usf.edu}
\affiliation{%
  \institution{University of South Florida}
  \city{Tampa}
  \state{Florida}
  \country{USA}
}

\author{Yiyao Xu}
\email{yiyaoxu@usf.edu}
\affiliation{%
  \institution{University of South Florida}
  \city{Tampa}
  \state{Florida}
  \country{USA}
}

\author{Chaoyun Yang}
\email{yangchaoyun@tongji.edu.cn}
\affiliation{%
  \institution{Tongji University}
  \city{Shanghai}
  \country{China}
}

\author{Lingyao Li}
\email{lingyaoli@arizona.edu}
\affiliation{%
  \institution{University of Arizona}
  \city{Tucson}
  \state{Arizona}
  \country{USA}
}

\author{Jingran Sun}
\email{jingransun@usf.edu}
\affiliation{%
  \institution{University of South Florida}
  \city{Tampa}
  \state{Florida}
  \country{USA}
}

\author{Hao Zhou}
\authornote{Corresponding author.}
\email{haozhou1@usf.edu}
\affiliation{%
  \institution{University of South Florida}
  \city{Tampa}
  \state{Florida}
  \country{USA}
}

\renewcommand{\shortauthors}{Wang et al.}

\begin{abstract}
Existing Level-2 driving-automation (DA) systems on production vehicles still rely on human drivers to decide when to engage automation, and ask for drivers' continuous attention and readiness to intervene in case of emergency.
This human--machine-interface (HMI) design demands good situational judgment and imposes high cognitive loads, producing a steep learning curve for new drivers, poor DA user experience, and possibly increased safety risks.
Improving DA HMIs hinges on accurately predicting when drivers hand control to automation and when they take it back, but no existing resource jointly captures both directions of driver--automation transitions with synchronized road, cabin, vehicle-control, and route observations at this scale.
To fill this gap, we introduce \textbf{\sysname}, a large-scale multimodal dataset of 781 real-world DA routes from 173 unique drivers across 108 vehicle models, spanning 204.9 hours of driving.
\sysname{} synchronizes front-view video, in-cabin video, decoded CAN (Controller Area Network) signals, radar-based lead-vehicle interaction, and GPS-derived route context into one record around each control transition.
The benchmark separates transition detection from anticipation: alongside handover prediction and an auxiliary action-recognition task, takeover is evaluated under two official protocols, onset detection (T3-D), where driver inputs are observable, and pre-override anticipation (T3-A), where every driver-override channel is withheld and windows end before the override begins.
We evaluate baselines spanning gradient-boosted trees, sequence models, cross-modal and hierarchical Transformers, and a frozen V-JEPA2 video encoder; zero-shot vision--language models are also benchmarked, all under a leakage-audited protocol.
Results show that i) multimodal context helps most on handover, where video world-model features raise AUPRC by 42\% over the strongest tabular baseline; ii) takeover onset detection is driven largely by observable driver-input cues; and iii) under the anticipation protocol all baselines score close to the base rate, indicating that the remaining signals carry little anticipatory information, establishing \sysname{} as a rigorous benchmark for multimodal driver--automation transition modeling.
\end{abstract}



\begin{teaserfigure}
  \centering
  \includegraphics[width=\textwidth]{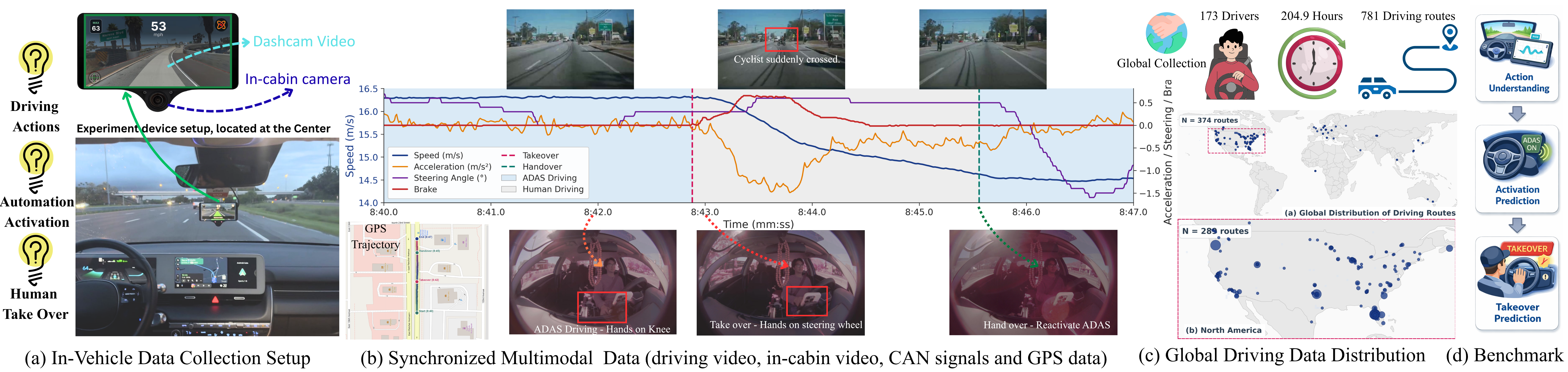}
  \caption{Overview of \textbf{\sysname}. Column (a) shows the data-collection setup with synchronized front-view and driver cameras; (b) the synchronized multimodal streams (road and cabin video, CAN signals, route context, lead-vehicle detections); (c) dataset scale and diversity (781 routes, 173 drivers, 204.9 hours); (d) the benchmark tasks.}
  \Description{
  A composite figure illustrating the \sysname~dataset: vehicle-mounted sensors for data collection, synchronized multimodal time-series signals with annotated control transition events, geographic distribution of driving routes, and benchmark tasks including action understanding, driving automation system (DAS) handover, and takeover prediction.
  }
  \label{fig:teaser}
\end{teaserfigure}

\begin{document}
\maketitle


\begin{table*}[htbp]
\begin{threeparttable}
\caption{Comparison with representative datasets and recent studies.}
\label{tab:dataset_comparison}
\centering
\small
\setlength{\tabcolsep}{3.6pt}
\renewcommand{\arraystretch}{1.00}
\begin{tabularx}{\textwidth}{
>{\raggedright\arraybackslash}p{1.8cm}
>{\raggedright\arraybackslash}p{1.5cm}
>{\raggedright\arraybackslash}p{3.2cm}
>{\raggedright\arraybackslash}p{2.4cm}
>{\raggedright\arraybackslash}p{2.4cm}
>{\raggedright\arraybackslash}X
}
\toprule
\lavrow Dataset & Setting & Modalities & Scale & Focus & Gap \\
\midrule
Drive\&Act~\cite{martin2019driveact}
& Controlled
& Cabin RGB/IR/depth
& 12 h
& Cabin actions
& No road view; no control transition \\

\rowcolor{lavA}
DAD~\cite{kopuklu2021dad}
& Simulator
& Cabin IR/depth
& 31 subjects
& Driver anomaly
& Simulator only; cabin only \\

AIDE~\cite{yang2023aide}
& Real-world
& Road + cabin video
& 2,898 clips
& Holistic Perception
& No control loop; not transition-centered \\

\rowcolor{lavA}
manD~\cite{nobari2024mand}
& Simulator
& Cabin + physiol. + vehicle
& 50 participants
& Driver Status
& Simulator only; not real-world driving \\

TD2D~\cite{hwang2025td2d}
& Simulator
& Cabin + takeover signals
& 500 cases; 50 drivers
& Takeover only
& Simulator only; one-sided transition \\

\rowcolor{lavA}
Lee \emph{et al.}~\cite{lee2025adas_activation}
& Real-world
& CAN + smartphone IMU
& 4 drivers
& Activation only
& Small scale; no cabin/road video \\

OpenLKA~\cite{wang2025openlka}
& Real-world
& Road video + CAN
& 400 h; 62 models
& LKA evaluation
& No cabin view; not interaction-centered \\

\rowcolor{lavA}
ADAS-TO~\cite{wang2026adas_to}
& Real-world
& Front-view + CAN
& 15,659 clips
& Takeover dataset
& No activation; no cabin view \\

\midrule
\rowcolor{lavB}
\textbf{\sysname\tnote{a}}
& \textbf{Real Daily Driving}
& \textbf{Road + cabin + radar + GPS + IMU + CAN}
& \textbf{204.9 h; 173 drivers; 781 routes}
& \textbf{Bidirectional transitions}
& \textbf{Real-world multimodal control-transition benchmark} \\
\bottomrule
\end{tabularx}

\begin{tablenotes}[flushleft]
\footnotesize
\item[a] \sysname\ adopts a similar collection methodology to OpenLKA and ADAS-TO, but contains no overlapping or reused data from either dataset.
\end{tablenotes}
\end{threeparttable}
\end{table*}

\section{Introduction}

Driving Automation (DA) systems are increasingly embedded in consumer vehicles, but today's advanced DA systems are not autonomous chauffeurs. NHTSA states that Level~2 systems can provide continuous assistance with both steering and acceleration/braking while the driver remains fully engaged, attentive, and responsible for the vehicle; its human-factors guidance further emphasizes that the driver must continuously monitor the roadway and be ready to intervene. Recent FIA Region~I findings likewise suggest that the safety benefits of DA depend not only on technical capability, but also on user engagement, satisfaction, acceptance, and trust. These facts make driver--automation control transitions a central problem in real-world assisted driving, i.e., drivers decide when to hand control to DA systems, and when to take it back \cite{nhtsa_driver_assistance, campbell2018l2l3_guidance, russell2021driver_expectations, fia_adas_dcas_2025}.

Studying this problem requires data that capture both sides of the transition together with the context surrounding it: the road scene outside the vehicle, the driver's state inside the cabin, the high-frequency vehicle control loop, interactions with leading vehicles, and route-level spatial context. However, existing data resources do not fully support this setting. Road-scene datasets mainly focus on external perception, driver-monitoring datasets often come from simulators or controlled laboratory studies, and takeover datasets are frequently one-sided or collected in controlled experimental settings. Representative examples include manD~1.0 for multimodal driver monitoring in a static simulator, TD2D for distracted takeover in an L2 simulator, ViE-Take for takeover under emotion-elicitation settings, and AIDE for assistive-driving perception with rich in-cabin and road-view signals but without bidirectional control transitions benchmarking as the primary task \cite{nobari2024mand, hwang2025td2d, wang2025vietake, yang2023aide, lee2025adas_activation}.

To address this gap, we present \sysname, a real-world multimodal benchmark for studying both when drivers hand control to the DA system and when they take it back. 
Our contributions are threefold:
i)~\textbf{Naturalistic multimodal dataset.}
We introduce \sysname{}, a real-world driving dataset spanning
781 routes, 173 drivers, 108 vehicle models, and 204.9 hours of
driving. The dataset
synchronizes front-view video, in-cabin video, CAN-decoded
vehicle dynamics, radar-based lead interaction, and GPS-derived
route context from diverse drivers, vehicles, and regions.
ii)~\textbf{A benchmark that separates detection from anticipation.}
On a frozen subset of 162.1 hours with 3,593 control-transition events, we define handover prediction, takeover \emph{onset detection} (T3-D), takeover \emph{pre-override anticipation} (T3-A, with every driver-override channel withheld and windows ending before the override begins), and an auxiliary signal-derived action-recognition task. The benchmark provides cross-driver and cross-vehicle splits, multiple horizons (1/3/5\,s), a leakage-audited input protocol, and sample-, event-, and operating-point (false-alarms-per-hour, warning lead time) metrics.
iii)~\textbf{Baselines and analysis.}
We evaluate gradient-boosted trees, sequence models, cross-modal and hierarchical Transformers, frozen video world-model features, and zero-shot vision--language models across single-modality and fusion settings. Multimodal context helps most on handover; takeover onset detection is largely driven by driver-input cues; and pre-override anticipation stays near the base rate for the same models that score highest on detection, suggesting that the remaining headroom reflects missing anticipatory information rather than limited model capacity. The benchmark package is released on \href{https://github.com/OpenLKA/BATON}{GitHub}, with a public benchmark tier that reproduces every reported result without any application, and the privacy-sensitive raw streams under identity-verified access on \href{https://huggingface.co/datasets/HenryYHW/BATON}{Hugging Face} (details in Appendix~\ref{app:release}).

\section{Related Work}

\subsection{Multimodal Driving and Behavior Datasets}
Existing datasets have advanced scene perception, driver monitoring, and in-cabin understanding, but offer limited support for studying real-world control transitions. Scene- and behavior-oriented datasets such as HDD, Drive\&Act, AIDE, and OpenLKA \cite{ramanishka2018hdd,martin2019driveact,yang2023aide,wang2025openlka} lack bidirectional handover coverage. Driver-focused datasets such as DAD \cite{kopuklu2021dad} and manD \cite{nobari2024mand} are simulator-based, while MDM \cite{jha2021mdm} provides a naturalistic multimodal corpus for driver attention rather than control-transition benchmarking. Real-world efforts such as AVDM \cite{sabry2024avdm} and ADABase \cite{oppelt2023adabase} do not jointly capture outside scene, driver state and vehicle control loop for transition analysis.

\subsection{Human--Automation Control Transitions}
Prior human-factors research has shown that control transitions are delayed, unstable, and shaped by traffic conditions, non-driving tasks, and driver state \cite{lu2016transitions,merat2014transition_manual,eriksson2017takeover_time,gold2016traffic_density,zhang2019determinants_takeover}, making handover and takeover central problems in transportation safety and HCI. Related multimodal modeling work has also examined takeover-side prediction, including DeepTake \cite{pakdamanian2021deeptake} and situational-awareness prediction during takeover transitions \cite{jia2024situational}. However, most existing datasets address only part of this problem: INAGT \cite{wu2021inagt} studies agent interaction timing rather than control transfer; TD2D and ViE-Take \cite{hwang2025td2d,wang2025vietake} focus on takeover in simulators; \citet{lee2025adas_activation} study real-world activation using only CAN and IMU from four drivers; and ADAS-TO \cite{wang2026adas_to} provides large-scale real-world takeover data but lacks activation events and in-cabin video. Driver-state resources from conditionally automated or simulated settings, such as the physiological takeover dataset of \citet{meteier2023physio} and the MPDB corpus~\cite{wang2024mpdb}, offer eye-tracking and physiological channels that \sysname{} does not measure; they are complementary, and richer driver-state sensing is one plausible route toward the pre-override anticipation problem our T3-A protocol leaves open. In contrast, \sysname{} supports real-world multimodal study of bidirectional control transitions (Table~\ref{tab:dataset_comparison}), synchronizing front-view video, in-cabin video, vehicle-control signals, radar interaction, and route context.

\section{The \sysname~Dataset}
\label{sec:dataset}

\begin{figure}[!htbp]
    \centering
    \includegraphics[width=0.8\linewidth]{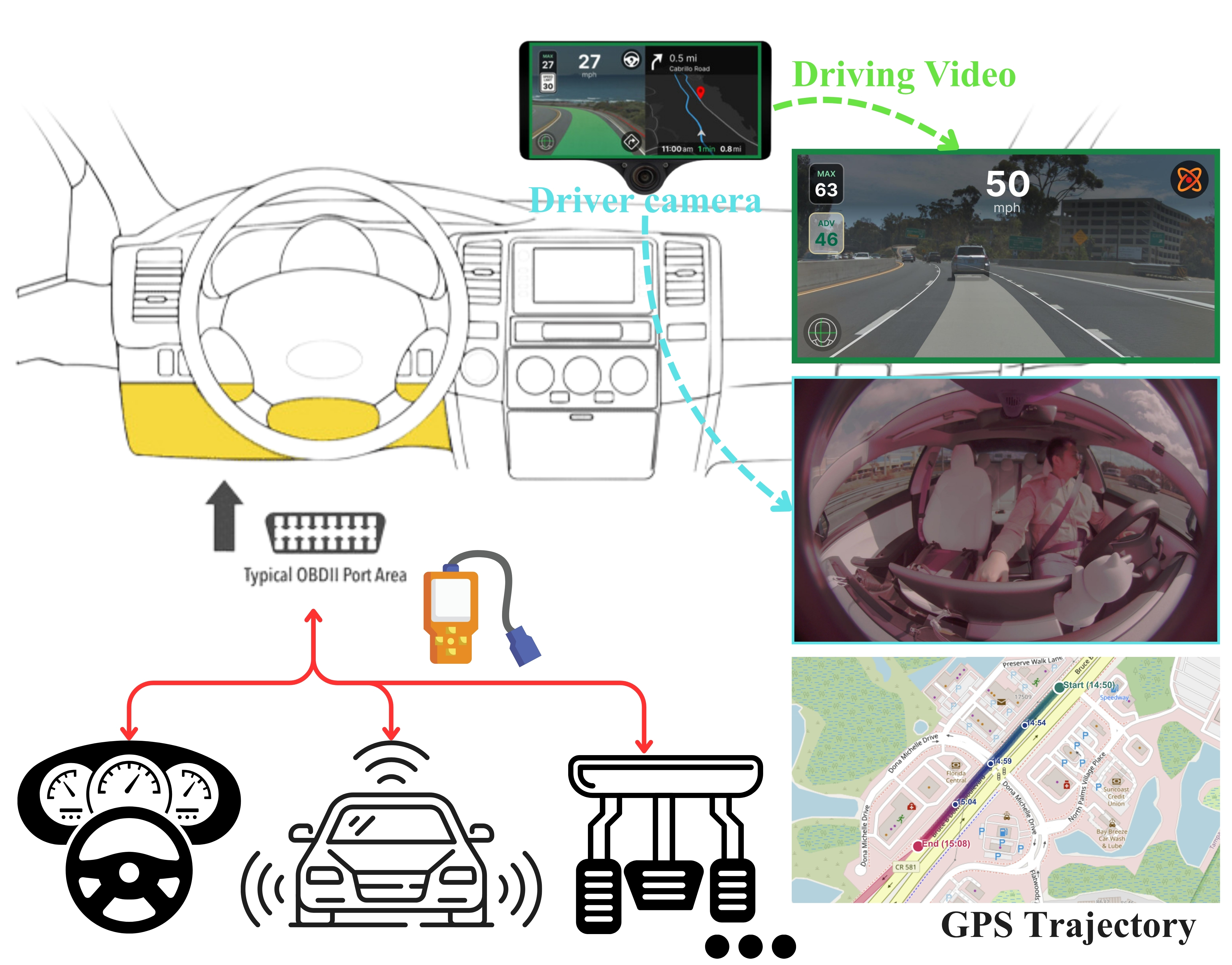}
    \caption{Data-collection setup. A comma~3X device~\cite{comma3x} records synchronized front-view and in-cabin video; CAN signals are decoded into vehicle-state measurements, and GPS provides route-level context.}
    \label{fig:collection_setup}
\end{figure}

\subsection{Dataset Collection Methods}

\sysname~is collected with comma devices mounted near the center of the front windshield, as illustrated in Fig.~\ref{fig:collection_setup}. This setup provides synchronized front-view and in-cabin video streams during everyday driving. In addition, we access vehicle CAN (Controller Area Network) signals through the onboard interface and decode them using Comma's public OpenDBC resources together with the cross-vehicle decoding pipeline released by OpenLKA \cite{wang2025openlka}. This allows us to recover fine-grained vehicle dynamics, control signals, and system states from a diverse set of production vehicles.

Our initial data collection is conducted in Tampa with five core drivers. We then expand the dataset geographically through direct collaboration, contributor outreach, and permission-based access to shared recordings. All recordings are contributed voluntarily by drivers who are aware of what their devices record; for every recording we obtained explicit permission for research use through direct communication with the contributor, independently of the fact that most of these routes had already been publicly shared by their owners on the comma/openpilot platform. A substantial part of our contribution therefore lies in this contributor communication, and in the curation, cleaning, synchronization, and organization that turn scattered contributed recordings into a coherent benchmark. This process substantially broadened the diversity of drivers, vehicles, and routes, enabling \sysname~to move beyond a small local collection and better reflect real-world human--automation driving across a wider range of environments.

\subsection{Data Processing}

After collection, raw route logs are converted into synchronized route-level signals, including vehicle dynamics, planning, radar, driver-state, IMU, GPS, and localization streams. GPS is then transformed into route-context features, including road type, speed limit, lane count, and proximity to intersections or ramps, while raw coordinates are excluded from benchmark inputs. The processed signals are used to define driving modes, detect handover and takeover events, generate driving-action labels, and construct benchmark samples and evaluation splits.

\subsection{Dataset Overview}

\sysname~is a real-world multimodal driving dataset built for studying bidirectional driver--automation control transitions. The current release contains 781 routes and 204.9 hours of driving from 173 drivers across 108 car models and 22 manufacturers, covering both human-driven and DA-assisted driving. Each driver in \sysname{} is a unique, pseudonymous driver ID: one recording setup is installed per vehicle and used by a single driver, so driver IDs correspond one-to-one to drivers and, in almost all cases, to their vehicles (only 4 of the 150 benchmark drivers recorded in more than one vehicle model; full driver--vehicle statistics in Appendix~\ref{app:dataset}). Of the 165.4 moving hours, 49.0\% are DA-engaged and 51.0\% are human-driven, so the dataset observes both control regimes in comparable volume. This scale and diversity make \sysname~suitable for a benchmark study of driver--DA interaction rather than a narrow case study.

\sysname{} is a living dataset: the corpus grew from 129 hours at its initial release in April 2026 to 204.9 hours within three months, and it continues to grow through ongoing community contributions. The benchmark is therefore defined on a frozen, quality-filtered subset: 565 route bundles totaling 162.1 hours from 150 drivers and 99 car models, in which our unified event definition identifies 3{,}593 control-transition events (1{,}800 DA handovers and 1{,}793 takeovers). All results in this paper are reported on this frozen subset; the remaining routes are released for future use.

\begin{figure}[!htbp]
    \centering
    \includegraphics[width=0.95\linewidth]{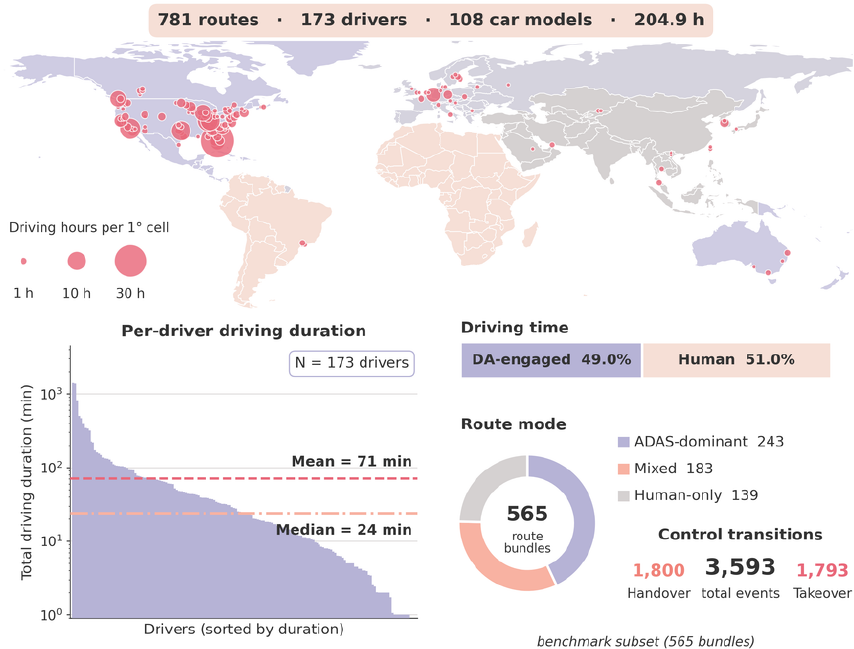}
    \caption{Overview of \sysname: global route distribution (top), driving time per driver (bottom left), and composition statistics (bottom right).}
    \label{fig:dataset_overview}
\end{figure}

\subsection{Modalities, Synchronization, and Coverage}

\sysname~provides synchronized multimodal observations of driver--ADAS interaction, including front-view video, in-cabin video, vehicle and control signals, radar-based lead interaction, driver-monitoring and planning signals, and GPS/localization context (Table~\ref{tab:modalities}). All modalities are aligned by their original logged timestamps at the route level. Coverage is high across the released dataset, with only a small number of routes missing GPS or front-view video; we retain these routes as part of a realistic real-world benchmark and document modality availability for filtering and task construction.

\begin{figure}[!htbp]
    \centering
    \includegraphics[width=\linewidth]{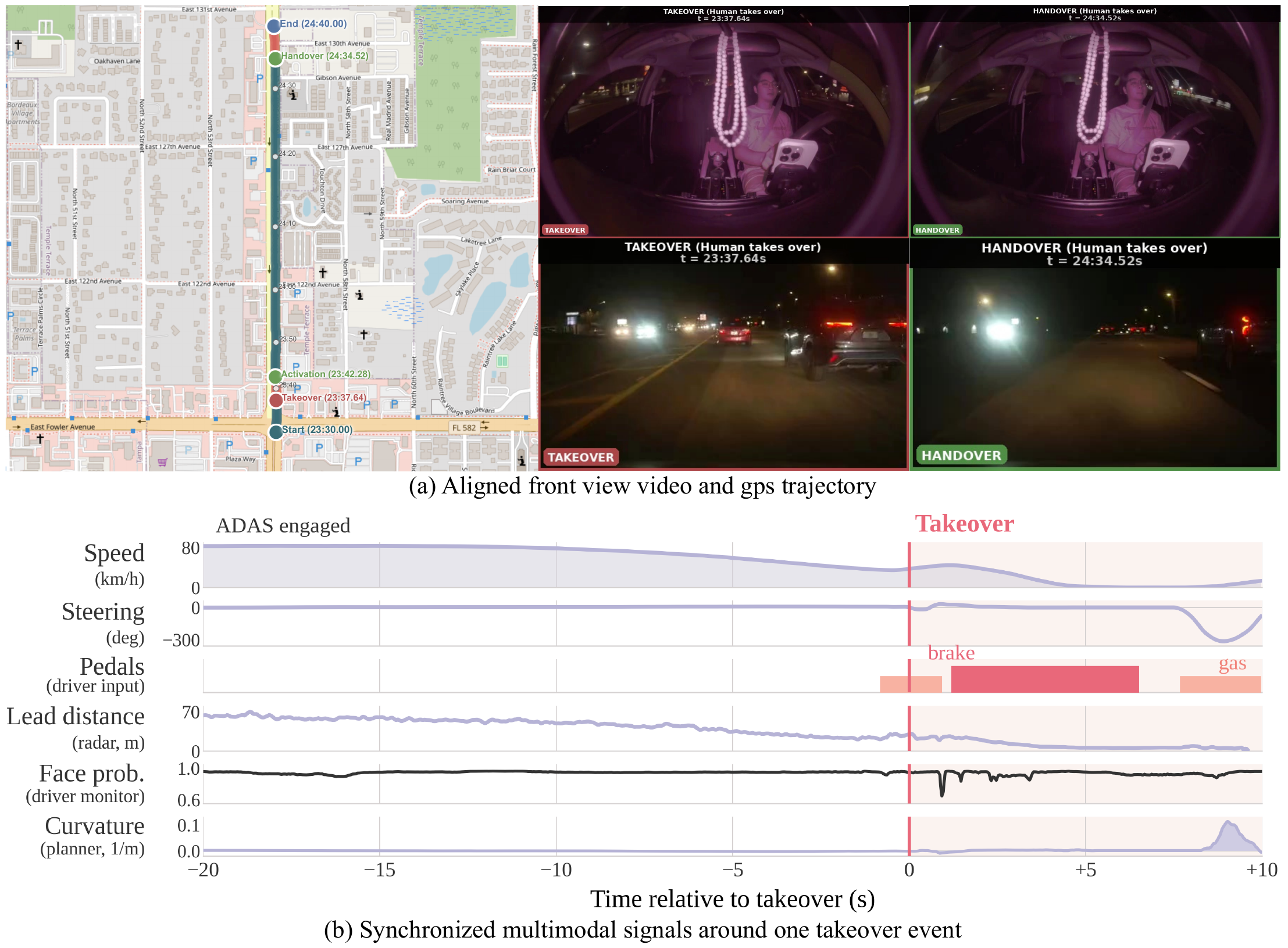}
    \caption{Multimodal context around control transitions in \sysname: (a) aligned cabin views, road views, and map context for takeover and handover events; (b) synchronized vehicle-dynamics, driver-input, radar, driver-monitoring, and planner signals around a single takeover event ($[-20\,\mathrm{s}, +10\,\mathrm{s}]$).}
    \label{fig:benchmark_overview_wide}
\end{figure}

\begin{table*}[htbp]
  \caption{Modalities in \sysname~and their roles in bidirectional control-transition analysis.}
  \label{tab:modalities}
  \centering
  \small
  \setlength{\tabcolsep}{3pt}
  \renewcommand{\arraystretch}{0.98}
  \begin{tabular*}{\textwidth}{@{\extracolsep{\fill}}
    >{\raggedright\arraybackslash}p{0.12\textwidth}
    >{\raggedright\arraybackslash}p{0.12\textwidth}
    >{\centering\arraybackslash}p{0.07\textwidth}
    >{\centering\arraybackslash}p{0.08\textwidth}
    >{\raggedright\arraybackslash}p{0.24\textwidth}
    >{\raggedright\arraybackslash}p{0.16\textwidth}
    >{\raggedright\arraybackslash}p{0.10\textwidth}
  }
    \toprule
    \lavrow Modality & Source & Rate & Coverage & Key parameters & Role & Data origin \\
    \midrule
    Front-view video  & Road camera   & 20 fps    & 777/781 & lanes, curves, traffic, lead vehicle & outside-scene context & raw video \\
    \rowcolor{lavA}
    In-cabin video    & Cabin camera  & 20 fps    & 781/781 & head pose, gaze, motion & driver readiness & raw video \\
    Vehicle dynamics  & CAN \& control & 100 Hz   & 781/781 & speed, steering, pedals, DA mode & control-loop state & CAN logs \\
    \rowcolor{lavA}
    IMU motion        & Device IMU    & 100 Hz    & 781/781 & acceleration, rotation & motion dynamics & inertial signals \\
    Radar interaction & Forward radar & 20 Hz     & 781/781 & relative distance, relative speed & lead interaction & radar tracks \\
    \rowcolor{lavA}
    Driver monitoring & DMS outputs   & 20 Hz     & 781/781 & awareness, distraction, eye state & driver state & openpilot \cite{commadriverattentionblog}  \\
    Planning state    & Planner outputs & 20 Hz   & 781/781 & target accel., warnings & assistance-stack output & CAN logs \\
    \rowcolor{lavA}
    GPS context       & GNSS / phone GPS & 10--20 Hz & $\sim$98\% & route, ramps, turns & spatial context & GNSS signals \\
    \bottomrule
  \end{tabular*}
\end{table*}


\subsection{Driving Modes and Control Transitions}

For benchmark construction, we define the driving mode by the automation-engagement state: we read directly from the CAN bus the flags that indicate whether an assistance stack (openpilot or the vehicle's stock ADAS) is actively controlling the car, and treat a segment as DA-active when either the lateral or the longitudinal flag is active, and as human-driven otherwise; these flags are used only to segment and filter driving modes, and their semantics are identical across stacks (Appendix~\ref{app:dataset} breaks the corpus down by assistance configuration; a single configuration accounts for 96\% of all transitions). Throughout the paper, \emph{handover} is shorthand for the manual-to-assisted transition (automation engagement) and \emph{takeover} for the assisted-to-manual transition (disengagement); under Level-2 assistance the driver retains supervisory responsibility at all times, so neither transition is a transfer of legal or full physical control. The binary state is a deliberate coarsening of a four-state lateral$\times$longitudinal taxonomy: lateral-only and longitudinal-only engagement together account for 16.3\% of driving time, and Appendix~\ref{app:dataset} reports the four-state time shares and per-subtype transition counts; subtype flags are released with the data. We keep the pooled binary target as the primary label because any assistance disengagement returns control authority to the driver and is therefore safety-relevant regardless of subtype, but we report subtype-stratified results so the pooled number cannot hide subtype imbalance. To suppress spurious toggles, we apply temporal filtering to remove short unstable episodes, retain only stable driving-state segments, and merge adjacent segments with the same stabilized state before extracting transitions (full filter parameters in Appendix~\ref{app:protocol}). Under the finalized benchmark protocol, 565 route bundles are retained, yielding 1{,}800 handover events and 1{,}793 takeovers.

\subsection{Release and Access}
\sysname{} is released in two tiers designed so that privacy protection never impairs benchmark reproducibility. The \emph{public benchmark tier} requires no application of any kind: structured tensors and visual features at multiple embedding granularities (clip- and window-level video embeddings, pose features, and 50\,Hz structured tensors), route metadata, action labels, official Task~1/2/3 sample-definition CSVs for all horizons and protocols, split files, evaluation scripts, and baseline code are public at \href{https://github.com/OpenLKA/BATON}{GitHub} and \href{https://huggingface.co/datasets/HenryYHW/BATON-Sample}{HuggingFace}; in-cabin content ships as rich, non-invertible embeddings, and this tier alone reproduces every number in this paper. The \emph{privacy-sensitive raw tier} covers raw video and per-route sensor streams. Most research needs are served without it: the public tier ships visual embeddings from multiple encoders, and we continue to add embedding types on request. For needs the embeddings cannot cover, our processing servers produce custom-format derived views under a reasonable-use request. Raw driver-facing video itself is released only under strict screening at \href{https://huggingface.co/datasets/HenryYHW/BATON}{HuggingFace}, and contributors can withdraw their data at any time.

\section{Benchmark Task Definition}
\label{sec:tasks}

Based on the driving modes and control-transition events defined above, \sysname~defines three benchmark tasks: (i) signal-derived action recognition, an auxiliary context task; (ii) handover prediction; and (iii) takeover, which we split into two official protocols, \emph{onset detection} (T3-D) and \emph{pre-override anticipation} (T3-A), because they measure different abilities. All tasks operate on synchronized multimodal observation windows under a unified protocol (Table~\ref{tab:benchmark_protocol}).

\begin{figure}[htbp]
    \centering
    \includegraphics[width=\linewidth]{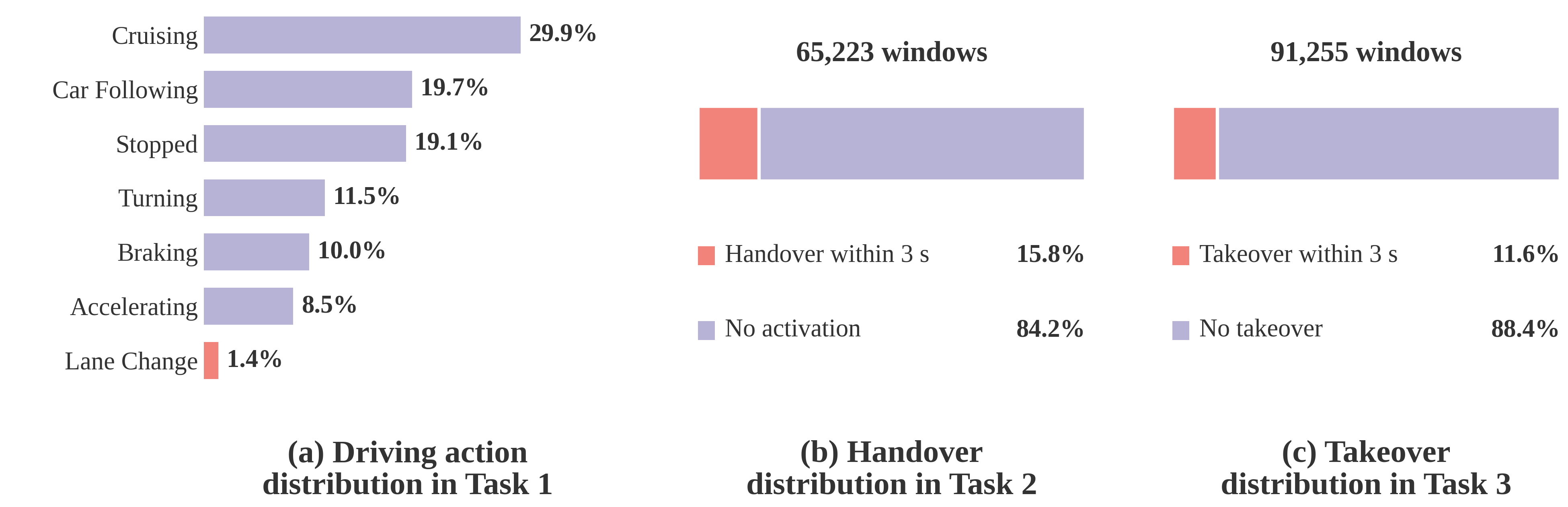}
    \caption{Distributions in the \sysname{} benchmark: (a) the seven Task-1 action classes; (b, c) positive and negative windows for Task~2 and Task~3.}
    \label{fig:task_distribution}
\end{figure}

\subsection{Task 1: Signal-Derived Action Recognition}

This auxiliary task provides short-term behavioral context for the two transition tasks; it is not a primary benchmark target. We formulate it as a coarse action recognition problem with seven classes: \textit{Cruising}, \textit{Accelerating}, \textit{Braking}, \textit{Turning}, \textit{Lane Change}, \textit{Stopped}, and \textit{Car Following} (Fig.~\ref{fig:task_distribution}(a)). Labels are assigned automatically from synchronized vehicle-state, planning, and lead-interaction signals using a rule-based protocol (rules in Appendix~\ref{app:protocol}), and each 5\,s sample is labeled by aggregating the per-second action labels within the window. Because the labels are a deterministic function of a subset of the signals, a model given those signals largely reconstructs the rules; we therefore define the \emph{official Task-1 evaluation on rule-free inputs} (modalities the rules never use), where a structured-signal model reaches 0.577 Macro-F1 and front-camera video alone reaches 0.592, against a seven-class chance level of 0.14. The full-signal setting (0.889) is kept only as a labeling sanity check in Appendix~\ref{app:vlm}. Label validity is supported by a threshold-sensitivity analysis and a blind human validation of 500 windows by a single domain expert (96.0\% agreement, Cohen's $\kappa=0.953$; a multi-annotator agreement study is left to future releases), detailed in Appendix~\ref{app:protocol}. Because the priority ordering collapses co-occurring actions into one class, we additionally release a \emph{multi-label} variant without the priority rule: 31.2\% of windows carry two or more concurrent action labels (e.g., braking-while-turning), and Appendix~\ref{app:vlm} reports the per-class label rates and multi-label baselines. For the single-label task we report Accuracy and Macro-F1.





\subsection{Task 2: Handover Prediction}

Task~2 predicts Human$\rightarrow$DA transitions. Given a 5\,s multimodal observation ending at time $t$ during manual driving, the model predicts whether the driver will activate DA within a future horizon $[t, t+h]$ (Fig.~\ref{fig:task_distribution}(b)). Samples are extracted at a 0.5\,s stride. Positive samples are constructed from pre-handover intervals, while negative samples are drawn from manual-driving intervals that remain transition-free around the prediction horizon. The benchmark provides 1\,s, 3\,s (main), and 5\,s horizon variants, containing 38{,}383, 65{,}223, and 76{,}366 samples, respectively. We report AUROC, AUPRC (primary), and F1.

\subsection{Task 3: Takeover Detection and Anticipation}

Task~3 concerns DA$\rightarrow$Human transitions. The sample construction mirrors Task~2: given a 5\,s multimodal observation ending at time $t$ during DA-active driving, the model predicts whether the driver will take back control within $[t, t+h]$ (Fig.~\ref{fig:task_distribution}(c)); the 1\,s, 3\,s, and 5\,s variants contain 48{,}293, 91{,}255, and 108{,}534 samples. A takeover, however, is physically initiated by the driver's own braking, accelerator, or steering input, and those signals are legitimately observable before the engagement flag switches. A model with access to them can therefore score well by recognizing an override \emph{already in progress}, which is not the same ability as anticipating the driver's intent. We make this distinction part of the benchmark by defining two official protocols:

\textbf{T3-D: takeover onset detection.} The leak-safe input set, which keeps the driver-input channels. This measures how early an in-progress transition can be recognized before the automation state changes, which is useful in itself (e.g., for smoothing the transfer of control) but is largely an early-detection problem.

\textbf{T3-A: pre-override takeover anticipation.} The anticipation-safe input set additionally withholds the six driver-override channels (brake, accelerator, steering pressure and torque), and every positive window must end at least 1\,s \emph{before} the driver's override onset. No part of the driver's takeover action is visible, so any score above the base rate reflects genuine anticipation from context (road scene, driver state, vehicle motion). Full protocol details are in Appendix~\ref{app:protocol}.

Because the label pools the engagement subtypes, the pooled target is best read as \emph{any-assistance} disengagement; full both-axes disengagement is reported separately in the subtype stratification (Appendix~\ref{app:vlm}). Task~2 does not need an analogous split: activating the automation is a button press whose surrounding driver inputs do not constitute the transition itself, and the flag-defining variables are already withheld by the leak-safe protocol.

\begin{table}[htbp]
\caption{\sysname~benchmark protocol.}
\label{tab:benchmark_protocol}
\centering
\small
\setlength{\tabcolsep}{4pt}
\renewcommand{\arraystretch}{1.05}
\begin{tabular*}{\columnwidth}{@{\extracolsep{\fill}}p{0.27\columnwidth}p{0.65\columnwidth}@{}}
\toprule
\lavrow Item & Setting \\
\midrule
Scope        & Bidirectional driver--automation transitions \\
\rowcolor{lavA}
Tasks        & T1 action recognition (auxiliary); T2 handover; T3-D takeover detection; T3-A takeover anticipation \\
Input / Horizon & 5\,s window; 1\,/\,3\,/\,5\,s horizon (main: 3\,s) \\
\rowcolor{lavA}
Stride       & 0.5\,s \\
Splits       & Cross-driver (main), cross-vehicle, random \\
\rowcolor{lavA}
Inputs       & Leak-safe (4 ADAS-control variables withheld); T3-A additionally withholds 6 driver-override channels \\
Seeds        & 3 seeds (42/123/7), mean $\pm$ std \\
\rowcolor{lavA}
Metrics      & T1: Accuracy, Macro-F1; T2/T3: AUPRC (sample- and event-level), AUROC, F1 \\
\bottomrule
\end{tabular*}
\end{table}



\subsection{Benchmark Splits and Evaluation Protocols}

We adopt cross-driver as the primary evaluation setting. Because each driver drives their own vehicle, holding out drivers also holds out their vehicles in almost all cases; the split therefore measures joint generalization to unseen drivers and their platforms, which we further disentangle with the cross-vehicle split (vehicle-model fingerprints held out), a within-driver temporal split, and train/test overlap slices (Appendix~\ref{app:vlm}). All split files are released. The finalized cross-driver split contains 405 routes for training, 84 for validation, and 76 for testing (104/26/20 drivers, disjoint across splits); a random split is also provided as a reference. Unless stated otherwise, all reported results use the cross-driver split, the 3\,s horizon, and three random seeds (42/123/7, mean $\pm$ std, no seed selection).

\textbf{Leak-safe inputs.} For the two prediction tasks the label is defined from the automation-engagement state, so we conservatively withhold the four ADAS-system control variables from all model inputs: the two flags that indicate whether lateral and longitudinal assistance are engaged, and the two automation-internal control signals, namely the controller's longitudinal state and its commanded acceleration (exact field names in Appendix~\ref{app:protocol}). Driver inputs (brake, accelerator, steering) are genuine real-time observations available at prediction time and are kept. Three checks indicate that no retained signal acts as a near-proxy for the label: the removed and retained field sets do not intersect, the strongest retained single feature reaches an AUPRC of 0.247 against 0.332 for the removed engagement flag, and a logistic reconstruction probe does not recover the label from the retained features (details in Appendix~\ref{app:protocol}).

\textbf{Event-level metric.} Because the 0.5\,s stride produces overlapping, correlated windows, we additionally report event-level AUPRC: the score of an event is the maximum over its windows, and negatives are non-overlapping horizon-length bins scored by their maximum. Sample-level and event-level results are reported side by side.

The benchmark package at \href{https://github.com/OpenLKA/BATON}{GitHub} contains the official sample-definition CSVs, split files, labels, generation code, evaluation scripts, and baseline code; the per-route sensor streams needed to build model inputs are hosted on HuggingFace. Appendix~\ref{app:release} gives the exact contents of each access tier and the steps to reproduce the results in this paper.

\begin{table}[htbp]
\caption{Protocol ladder (single XGBoost, cross-driver, $h{=}3$\,s, sample\,/\,event AUPRC). Base rates: T2 0.148\,/\,0.037; T3 0.120 leak-safe, 0.035 anticipation-safe.}
\label{tab:leakage}
\centering
\small
\setlength{\tabcolsep}{4pt}
\renewcommand{\arraystretch}{1.05}
\begin{tabular*}{\columnwidth}{@{\extracolsep{\fill}}p{0.44\columnwidth}p{0.22\columnwidth}p{0.22\columnwidth}@{}}
\toprule
\correow Input protocol & T2 & T3 \\
\midrule
Full input (upper-bound reference) & 0.360 / 0.292 & 0.520 / 0.428 \\
\rowcolor{corLight}
Leak-safe (official; T3-D) & 0.222 / 0.091 & 0.478 / 0.385 \\
Anticipation-safe (T3-A) & -- & 0.056 / 0.026 \\
\bottomrule
\end{tabular*}
\end{table}

\begin{table*}[htbp]
\caption{Main transition results (cross-driver, $h{=}3$\,s, sample\,/\,event AUPRC, 3-seed mean$\pm$std; T2 and T3-D leak-safe, T3-A anticipation-safe with base rate 0.035). $^{s}$:~single seed.}
\label{tab:main_results}
\centering
\small
\setlength{\tabcolsep}{4pt}
\renewcommand{\arraystretch}{1.05}
\begin{tabular*}{\textwidth}{@{\extracolsep{\fill}} ll ccc}
\toprule
\correow Method & Input & T2: Handover & T3-D: Takeover detection & T3-A: Takeover anticipation \\
\midrule
Base rate & -- & .148 / .037 & .120 / -- & .035 / -- \\
GRU & struct & .202$\pm$.009 / .072$\pm$.006 & .280$\pm$.049 / .119$\pm$.029 & -- \\
Cross-Modal Transformer & struct & .242$\pm$.011 / .103$\pm$.005 & .316$\pm$.028 / .135$\pm$.031 & -- \\
RG-HBT-Q & struct+video & .254$\pm$.020 / .111$\pm$.020 & .398$\pm$.007 / .230$\pm$.006 & .058$\pm$.006 / .026$\pm$.002 \\
DI-RG-HBT-Q & struct+video & .219$\pm$.011 / .116$\pm$.017 & .366$\pm$.025 / .215$\pm$.036 & -- \\
TCN & struct & .190$\pm$.008 / .058$\pm$.008 & .332$\pm$.064 / .167$\pm$.046 & -- \\
V-JEPA2 fusion & video+struct & .235$\pm$.005 / .089$\pm$.004 & .329$\pm$.037 / .161$\pm$.032 & -- \\
XGBoost & CAN statistics & .236$\pm$.003 / .103$\pm$.008 & .479$\pm$.007 / .380$\pm$.002 & .056$\pm$.001 / .026$\pm$.001 \\
\midrule
\rowcolor{corLight}
\textbf{\sysname-WM} & multimodal & \textbf{.335$\pm$.003} / \textbf{.171$\pm$.005} & \textbf{.514$\pm$.000} / \textbf{.413$\pm$.003} & .070$\pm$.001 / .040$\pm$.004 \\
\bottomrule
\end{tabular*}
\end{table*}

\begin{table}[htbp]
\caption{Zero-shot VLM baselines (cross-driver, $h{=}3$\,s, sample-level, best configuration per task, identical frozen sample draw). T1 uses rule-free camera-only inputs; the Full configuration contains rule-defining vehicle-state text and appears in Appendix~\ref{app:vlm} as a leakage-inflated diagnostic.}
\label{tab:vlm_results}
\centering
\small
\setlength{\tabcolsep}{4pt}
\renewcommand{\arraystretch}{1.05}
\begin{tabular*}{\columnwidth}{@{\extracolsep{\fill}}p{0.42\columnwidth}ccc@{}}
\toprule
\correow Model & T1 $F1_M$ & T2 & T3-D \\
\midrule
Gemini-2.5-Flash & 0.362 & 0.184 & 0.202 \\
GPT-4o & 0.303 & 0.165 & 0.196 \\
Qwen3-VL-8B & 0.280 & 0.166 & 0.177 \\
Qwen3-VL-4B & 0.212 & 0.173 & 0.167 \\
MiniCPM-V-4.5 & 0.170 & 0.174 & 0.182 \\
LLaVA-OneVision-1.5-8B & 0.168 & 0.195 & 0.149 \\
InternVL3.5-8B & 0.142 & 0.159 & 0.161 \\
\bottomrule
\end{tabular*}
\end{table}

\section{Experiments}
\label{sec:baselines}

We evaluate \sysname{} under one official protocol: frozen benchmark subset, leak-safe inputs, cross-driver split, $h{=}3$\,s, three seeds (mean$\pm$std), and sample- and event-level AUPRC. The baselines cover sequence and fusion models (GRU, a cross-modal Transformer, and the hierarchical RG-HBT-Q / DI-RG-HBT-Q models with modality dropout), gradient-boosted trees (XGBoost~\cite{chen2016xgboost}) on per-channel window statistics, frozen V-JEPA2 video features~\cite{assran2025vjepa2}, a video world-model readout, and zero-shot VLMs (10 open models, e.g., Qwen-VL~\cite{bai2025qwen25vl}, plus Gemini and GPT-4o~\cite{openai2024gpt4o}). Structured signals are resampled to 50\,Hz. Architectures, training configurations, and hardware are in Appendix~\ref{app:baselines}.

\subsection{The Leakage-Avoidance Protocol}

Table~\ref{tab:leakage} traces one model down the input-protocol ladder. Withholding the four ADAS-control variables (Section~\ref{sec:tasks}) removes direct label leakage and costs T2 most of its apparent headroom, confirming that those variables carried label-correlated signal; the leak-safe level defines T2 and T3-D. Descending to the anticipation-safe level, which withholds the six driver-override channels and ends every positive window at least 1\,s before the override begins, collapses T3 from 0.478 to 0.056 AUPRC. Nearly all of the takeover-detection performance therefore comes from observing the driver's action itself. The collapse is unlikely to reflect limited model capacity: the identical feature sets and readouts reach 0.479--0.514 on T3-D, the strongest additions in our suite (video world-model features, driver pose, hierarchical multimodal fusion) all help on other tasks, and none of them lifts T3-A meaningfully above the 0.035 base rate (Table~\ref{tab:main_results}). Longer context does not change the picture either: extending the observation window to 30--60\,s, attending over V-JEPA2 video, pose, and structured streams with a cross-modal transformer, distilling from the strongest tabular teacher, and privileged-information distillation from a detection teacher that sees the override during training all stay within 0.033--0.070 (Appendix~\ref{app:vlm}). The retained signals appear to carry little anticipatory information, leaving richer driver-state sensing, such as gaze or physiology, as the more plausible path forward. The gap holds up statistically despite the smaller anticipation-safe test set (92 events). Event-level bootstrap 95\% CIs for detection (0.31--0.44) and anticipation (0.019--0.041) are disjoint by an order of magnitude. The driver-disjoint validation split provides a second, larger set of 160 anticipation events, and the picture is the same there: anticipation stays near its 0.039 base rate (0.075--0.102 sample AUPRC, and this split was used for hyperparameter selection, which would bias it upward if anything) while detection reaches 0.52. Relaxing the pre-override buffer to 0.5\,s enlarges the test set to 96 events with the same result, and the collapse decays smoothly as the buffer varies from 0.5\,s to 2.0\,s (Appendix~\ref{app:vlm}); 92.5\% of benchmark takeovers are driver-initiated, so the filtered positive class is not dominated by heterogeneous system-initiated events. Pre-override takeover anticipation is the open problem that \sysname{} makes measurable.

\subsection{Main Comparison}

Table~\ref{tab:main_results} compares baselines under the official protocol. On T2, the V-JEPA2 fusion model reaches 0.235 and features derived from video alone reach 0.286 (Appendix~\ref{app:vlm}), which shows that video carries an independent predictive signal; augmenting the tabular baseline with video world-model features raises it from 0.236 to 0.335 (+42\%). On T3-D the ordering reverses: the CAN-statistics XGBoost (0.479) clearly beats every neural multimodal model (best 0.398), consistent with the well-documented advantage of tree-based models on tabular data at this sample scale~\cite{grinsztajn2022tabular}, and adding driver-pose features lifts it to 0.514. Under the current protocol, then, the observed predictive cues differ by direction: scene and driver-state context contributes most where the transition is not preceded by a sharp driver action (handover), while takeover detection is dominated by driver-input dynamics that tree-based models capture well. The per-task feature composition of \sysname-WM (world-model features for T2, pose features for T3) is itself evidence that no single fusion recipe currently wins across tasks. Stratifying test events by engagement subtype further shows that current performance concentrates on lateral-only transitions (T3-D event AUPRC 0.72 for lateral-only vs.\ 0.05 for simultaneous lateral+longitudinal takeovers; Appendix~\ref{app:vlm}), so the pooled binary target understates how uneven the solved and unsolved parts of the problem are.

\subsection{False-Alarm and Lead-Time Evaluation}

Event-level AUPRC still credits detections that arrive arbitrarily late within the horizon. We therefore also evaluate deployment-style operating points: sweeping the decision threshold to a fixed false-alarm budget per driving hour and asking how many events are caught, and how early (Appendix~\ref{app:vlm}). At 1 false alarm per hour, the best T3-D model recalls only 28\% of takeover events (median earliest warning 3.0\,s, the horizon cap); the best T2 model recalls 7\% of handovers (world-model features roughly double the tabular baseline's 3.7\%); and T3-A recall is at most 2\%. These numbers are far below deployment needs and complement the threshold-free metrics.

\subsection{Comparison with a Deployed Predictor}

The driving model deployed in openpilot predicts its own disengagement probabilities at five horizons, which makes it a natural practical reference. We recovered these predictions from the original vehicle logs, available for 26 benchmark routes (5 in the test split), and scored each window by its maximum predicted disengagement probability. On this covered subset the deployed predictor reaches 0.293\,/\,0.101 sample/event AUPRC on T3-D, above the 0.122 base rate but far below our tabular baseline on the same rows (0.771\,/\,0.732); the anticipation-safe subset is too small to evaluate reliably (9 positives). A production system's own takeover predictor therefore leaves most of the task unsolved, which supports the need for a dedicated benchmark (details in Appendix~\ref{app:vlm}).

\subsection{Generalization Axes}

We disentangle what the primary split measures with three probes (full tables in Appendix~\ref{app:vlm}). (i) \emph{Cross-vehicle} (vehicle-model fingerprints held out): on T2 every method scores higher than under cross-driver (e.g., XGBoost 0.360 vs.\ 0.236), and on T3-D the neural models do as well, so unseen drivers (each with their own vehicle) are the harder axis for handover; T3-D tabular models are the one exception (0.443--0.452 vs.\ 0.479--0.514). (ii) \emph{Within-driver temporal} (same drivers, later routes held out): T2 rises to 0.454--0.465, a +0.19--0.22 lift over cross-driver, showing that most of the cross-driver gap on handover comes from identity (behavior, platform, controller semantics) rather than temporal drift; T3-D is far less identity-sensitive (+0.05). (iii) \emph{Familiarity slices} of the cross-driver test set: the seen-vs-unseen vehicle-model gap flips sign between tasks and every software-version series in test also appears in training, so neither factor systematically drives the reported numbers. World-model and distilled-fusion methods are excluded from (i)--(ii) because their pretraining is tied to the cross-driver training routes.

\subsection{Zero-Shot VLMs}

All twelve VLMs are evaluated on one frozen sample draw with identical prompts and parsing (Table~\ref{tab:vlm_results}). The pattern is consistent: performance is highest with front-view and text-context inputs combined, cabin-only input is near chance, and even the strongest closed model stays far below trained baselines on T2/T3-D, indicating that recognizing control transitions from a handful of sampled frames is difficult for current VLMs. Under rule-free inputs the closed models' advantage over the best open model (Qwen3-VL-8B) persists on Task~1 (0.362 vs.\ 0.280) but nearly vanishes on the transition tasks.

\section{Discussion}
\label{sec:discussion}

\sysname{} provides a unified benchmark for bidirectional driver--automation control transitions in naturalistic driving. The baseline results show that selected feature augmentations help selected tasks under this protocol: video world-model features raise handover prediction substantially, driver-pose features raise takeover detection, and no single fusion recipe wins across tasks. This suggests that road context, driver state, and vehicle dynamics provide complementary but task-dependent cues. The gap between current results and practical performance, most visible at deployment-style operating points, suggests substantial room for stronger multimodal architectures. Under the current protocol, the two transition directions expose different predictive cues: scene and driver-state context matters most for handover, whereas takeover onset detection is dominated by driver-input dynamics. Once those inputs are withheld (T3-A), the same models that score highest on detection stay close to the base rate, which suggests an information limit of the current signal set. The detection/anticipation split quantifies this headroom, and richer driver-state sensing appears the most plausible route to closing it.

\textbf{Limitations.}
First, \sysname{} provides front-view observations only, without BEV-style surrounding-vehicle context. Second, driving time is unevenly distributed across drivers (the top 5 of 150 drivers account for 31\% of benchmark hours; 80 drivers contribute under 30 minutes). Third, Task-1 labels are rule-derived and validated by a single annotator; the priority ordering also collapses co-occurring actions into one class. Fourth, the released baselines rely on relatively simple fusion and the strongest reference model uses task-specific feature sets rather than a unified architecture.

\textbf{Future work.}
Future work will expand driver, route, and vehicle diversity, incorporate richer surrounding-context representations, and develop stronger multimodal and personalized models for control-transition prediction.

In summary, \sysname{} provides synchronized multimodal data and benchmark tasks for studying driver--automation control transitions in real-world driving.

\section{Ethical Considerations and Privacy}
All data in \sysname\ were contributed voluntarily by the drivers themselves, and for every recording we obtained explicit research-use permission through direct communication with the contributor; we do not treat public platform sharing as a substitute for this permission. Our role centered on this communication and on curating and organizing the recordings. All data were collected and processed in accordance with applicable privacy requirements, participant-consent procedures, and platform terms where applicable. For recordings contributed from the comma/openpilot ecosystem, collection context follows comma's publicly posted Terms and Privacy Policy~\cite{comma_terms_privacy} and contributor permission. To reduce privacy risks, raw GPS coordinates are removed from the benchmark and replaced with semantically derived route-context features, directly identifying information is removed from vehicle logs, sensitive visual content is anonymized, and the public tier distributes in-cabin content as non-invertible embeddings; in particular, all occupants inside the vehicle cabin other than the driver have their faces blurred. This work is a secondary use of platform-collected data under the platform's terms and explicit contributor permission; contributors can request removal of their routes at any time, removals propagate to every tier, and roadside faces and license plates in publicly distributed video are automatically detected and blurred with human spot-check auditing. Appendix~\ref{app:release} details the access tiers, consent basis, withdrawal process, and privacy-risk assessment.

\begin{acks}
We sincerely thank all drivers and driving-automation enthusiasts who voluntarily contributed data to this project. Their participation and support were essential to the collection and release of this dataset and benchmark.
\end{acks}

\newpage

\bibliographystyle{ACM-Reference-Format}
\bibliography{references}

\clearpage
\appendix

\section{Dataset Details}
\label{app:dataset}

\subsection{Collection Hardware and Logging}
All routes are recorded with comma devices (comma three / comma 3X) mounted at the center of the front windshield. The device records a forward-facing road camera and an in-cabin camera at 20\,fps, an on-device IMU at 100\,Hz, and GNSS position. Through the car harness, the device reads the vehicle CAN buses; openpilot's logging stack stores all streams with their original logged timestamps on a common route-level clock, so no post-hoc cross-sensor alignment is required. CAN frames are decoded into physical signals with comma's public OpenDBC definitions together with the cross-vehicle decoding pipeline of OpenLKA~\cite{wang2025openlka}, which lets one consistent signal schema cover the 108 vehicle models in the corpus. Recordings are uploaded route-by-route; a route corresponds to one contiguous drive segment.

\subsection{Corpus vs.\ Benchmark Subset}
\sysname{} is maintained as a living dataset: new contributor routes continue to be ingested through the same pipeline. To keep results comparable, all numbers in this paper are computed on a frozen benchmark subset (internally \texttt{benchmark\_v2}) that passes quality filters on signal completeness, video decodability, and minimum route length. Table~\ref{tab:corpus_stats} summarizes both scopes.

\begin{table}[htbp]
\caption{Full living corpus vs.\ the frozen benchmark subset used for all reported results.}
\label{tab:corpus_stats}
\centering
\small
\setlength{\tabcolsep}{4pt}
\renewcommand{\arraystretch}{1.05}
\begin{tabular*}{\columnwidth}{@{\extracolsep{\fill}}p{0.44\columnwidth}p{0.24\columnwidth}p{0.24\columnwidth}@{}}
\toprule
\neurow Statistic & Full corpus & Benchmark subset \\
\midrule
Routes / route bundles & 781 & 565 \\
Driving hours & 204.9 & 162.1 \\
Drivers & 173 & 150 \\
Vehicle models (manufacturers) & 108 (22) & 99 \\
Control transitions & -- & 3{,}593 \\
\quad handover / takeover & -- & 1{,}800 / 1{,}793 \\
DA-engaged share of moving time & 49.0\% & -- \\
\bottomrule
\end{tabular*}
\end{table}

\subsection{Synchronization and Coverage}
All modalities keep their native rates (Table~\ref{tab:modalities}); the benchmark loader resamples structured signals to 50\,Hz and reads video at 2\,fps for feature extraction. Coverage over the full corpus: front-view video 777/781 routes, in-cabin video 781/781, vehicle dynamics / IMU / radar / driver-monitoring / planning 781/781, GPS $\sim$98\%. Routes with missing GPS or front video are retained in the corpus and documented, and are excluded from tasks that require the missing modality.

\subsection{Driver--Vehicle Structure}
Each of the 150 benchmark drivers has a unique, pseudonymous driver ID and drives their own vehicle. The 150 drivers span 99 vehicle-model fingerprints: 4 drivers recorded in more than one vehicle model, and 28 model strings are shared by two or more drivers (at most 7). Consequently, the cross-driver split is simultaneously a driver-level and (almost everywhere) a vehicle-level hold-out, while the cross-vehicle split holds out model fingerprints.

\subsection{Assistance Configurations and Flag Semantics}
The two engagement flags are read directly from the CAN bus and have stack-invariant semantics: the lateral flag is set whenever \emph{any} lane-keeping controller (openpilot or the vehicle's stock system) is actively steering, and the cruise flag whenever \emph{any} adaptive cruise control is engaged. Segmenting driving modes with these flags therefore does not depend on which stack is running. Classifying each benchmark route by the co-occurrence of these flags over its assisted time (Table~\ref{tab:assistcfg}) shows that the corpus is far more homogeneous than a mixed-stack setting: 96\% of all transition events come from a single configuration, openpilot lateral control combined with the vehicle's stock adaptive cruise, openpilot longitudinal control is absent, and purely stock assistance contributes under 4\% of events. Label semantics are therefore not materially confounded by stack diversity.

\begin{table}[htbp]
\caption{Benchmark routes by assistance configuration (classified from CAN engagement flags over assisted driving time).}
\label{tab:assistcfg}
\centering
\small
\setlength{\tabcolsep}{3.5pt}
\renewcommand{\arraystretch}{1.05}
\begin{tabular*}{\columnwidth}{@{\extracolsep{\fill}} l rrrr}
\toprule
\neurow Configuration & Routes & Hours & Handovers & Takeovers \\
\midrule
openpilot lateral + stock ACC & 409 & 128.0 & 1{,}728 & 1{,}726 \\
Stock assistance only & 31 & 11.7 & 69 & 63 \\
openpilot lateral + longitudinal & 0 & -- & -- & -- \\
Human-only routes ($<$10\,s assisted) & 125 & 22.4 & 3 & 4 \\
\bottomrule
\end{tabular*}
\end{table}

\subsection{Representativeness and Known Skews}
\textbf{Contribution imbalance.} The top 5 of 150 drivers account for 31.3\% of benchmark driving hours (top 20: 60.1\%); the median driver contributes 0.41\,h, the largest 22.7\,h, and 80 drivers contribute under 30 minutes. The top 5 vehicle models account for 40.8\% of hours. \textbf{Software and hardware.} Recordings span openpilot 0.7--0.11.x and comma-release builds (most frequent: 2026.001, 83 routes), on three recording-hardware generations (tizi 281, tici 148, mici 132 routes). \textbf{Time of day.} For the 221 routes (57.9\,h) with both a GPS fix and a timestamp, approximate local solar time (UTC offset by longitude) places 42\% of hours in daytime (07--19\,h) and 58\% at night; the remaining routes cannot be localized in time and are unlabeled. \textbf{Cabin-perception coverage.} Driver-monitoring outputs have no fully-missing frames on either day or night routes, and full-route pose extraction shows a near-zero all-zero row fraction in both conditions, so cabin-based features do not silently drop out at night at the frame level. \textbf{What is not covered.} No administrative region, weather, or demographic attributes are collected; fairness across such groups is therefore not evaluable from the released data.

\begin{table}[htbp]
\caption{Lateral$\times$longitudinal engagement structure of the benchmark subset. Subtype counts are raw flag changes before the debounce/merge filters of Appendix~\ref{app:protocol}.}
\label{tab:fourstate}
\centering
\small
\setlength{\tabcolsep}{4pt}
\renewcommand{\arraystretch}{1.05}
\begin{tabular*}{\columnwidth}{@{\extracolsep{\fill}}l rr@{}}
\toprule
\neurow Engagement state & Hours & Share \\
\midrule
Neither active (manual) & 73.3 & 45.2\% \\
Longitudinal only (ACC-style) & 11.0 & 6.8\% \\
Lateral only & 15.4 & 9.5\% \\
Both active & 62.4 & 38.5\% \\
\midrule
\neurow Raw OR-transition subtype & Activations & Deactivations \\
\midrule
Lateral flag only & 1{,}112 & 1{,}003 \\
Longitudinal flag only & 416 & 977 \\
Both flags together & 706 & 244 \\
\bottomrule
\end{tabular*}
\end{table}

\section{Release, Licensing, and Usage Norms}
\label{app:release}

\subsection{Access Tiers and Reproducibility}
Table~\ref{tab:tiers} states exactly what each tier contains and what it suffices to reproduce. The public benchmark tier ships the precomputed 50\,Hz structured tensors, video features, and pose features for all 565 benchmark routes, which suffice to reproduce every number in Tables~\ref{tab:leakage}--\ref{tab:vlm_results} without any application; in-cabin content is present only as non-invertible visual features, so restricted raw cabin video is never needed to use the benchmark. The privacy-sensitive raw tier is governed in three modes: continuously expanded public embeddings for most uses, server-side processed derived views under reasonable-use requests, and strictly screened access for raw driver-facing video; contributors can withdraw their data at any time and withdrawals propagate to every tier. The dataset is licensed CC~BY-NC~4.0; a LICENSE file, SHA-256 checksums for all sample-definition CSVs and split files, and a pinned reproduction command sequence (download $\rightarrow$ preprocess $\rightarrow$ train $\rightarrow$ evaluate) ship in the repository, and an archival DOI will be minted for the camera-ready version.

\begin{table}[htbp]
\caption{Release tiers and what each tier alone reproduces.}
\label{tab:tiers}
\centering
\footnotesize
\setlength{\tabcolsep}{3.5pt}
\renewcommand{\arraystretch}{1.1}
\begin{tabular*}{\columnwidth}{@{\extracolsep{\fill}}p{0.20\columnwidth}p{0.13\columnwidth}p{0.33\columnwidth}p{0.20\columnwidth}@{}}
\toprule
\neurow Tier & Access & Contents & Reproduces \\
\midrule
GitHub benchmark package & public & sample CSVs (all tasks/horizons incl.\ T3-A), split files, labels, event lists, generation/eval/baseline code & protocol, metrics, splits \\
\rowcolor{lavB!30}
HF \sysname-Sample & public & 43 full routes (video + all sensor CSVs) & pipeline inspection \\
Benchmark-ready tier & public & 50\,Hz tensors, video/pose features for 565 benchmark routes & all reported results \\
\rowcolor{lavB!30}
HF \sysname{} full corpus & strict screening & all 781 routes: video + per-route sensor CSVs, in full & everything, incl.\ new tasks \\
\bottomrule
\end{tabular*}
\end{table}

\subsection{Consent, Review Status, and Withdrawal}
Recordings originate from the comma/openpilot ecosystem under its publicly posted Terms and Privacy Policy~\cite{comma_terms_privacy}. Every route was contributed voluntarily by a driver who is aware of what the device records. The consent basis is uniform across the corpus: explicit permission for research use, obtained through direct communication with each contributor; the fact that the majority of routes had already been publicly shared by their owners on the platform provides context but is never treated as research consent. \sysname{}'s contribution for those recordings is accordingly communication, verification, curation, and standardization rather than new collection. This work is a secondary use of platform-collected data. No demographic attributes are collected, and no biometric templates are derived or released. Contributors can request removal of their routes through the repository contact; removals propagate to every tier and are recorded in the changelog of the next dataset version, and the identity-confirmation form keeps all source-data access traceable and accountable. Faces of all cabin occupants other than the driver are blurred; the driver-facing raw video is part of the identity-verified full release, while the zero-step public tier carries it only as non-invertible embeddings. Roadway-visible faces and license plates in publicly distributed front-camera video pass an automated detection-and-blurring pipeline with human spot-check auditing before release, and any missed instance reported through the repository is patched in the next versioned release. A privacy-risk assessment covering video-based re-identification and location inference from route imagery accompanies the release: raw GPS is withheld, route-context features are semantically abstracted, cabin content is feature-only in the zero-step public tier, and the identity-confirmation form keeps source-data access fully traceable.

\subsection{Privacy and De-identification}
Collection follows comma's publicly posted Terms and Privacy Policy~\cite{comma_terms_privacy} and explicit contributor permission. Before release: raw GPS coordinates are removed from all benchmark inputs and replaced with semantically derived route-context features (road type, speed limit, lane count, proximity to ramps and intersections); directly identifying fields (e.g., device and vehicle identifiers) are stripped from the logs; faces of all cabin occupants other than the driver are blurred; and remaining sensitive raw visual content is part of the identity-verified full release. We do not release any biometric templates, and the driver-monitoring stream contains only derived state outputs (attention, eye state), not raw face crops.

\subsection{Maintenance and Versioning}
The benchmark subset is frozen and versioned; future corpus growth will be released as new, additive versions with changelogs, and previously published splits will never be modified. Errata (e.g., corrected labels) will be released as clearly versioned patches, and issues can be reported through the public repository.

\subsection{Usage Norms}
A complete datasheet (Gebru et al.\ style: motivation, composition, collection, preprocessing, uses, distribution, maintenance), including contributor-recruitment description, known skews, and explicit guidance on when \sysname{} should not be treated as representative of general driver populations, ships as \texttt{DATASHEET.md} in the repository. The dataset is intended for research on driver--automation interaction, driving safety, and multimodal modeling. Users must not attempt to re-identify drivers, vehicles, or locations, must not use the data for surveillance or for scoring identifiable individuals, and should report the benchmark version and input protocol (leak-safe or anticipation-safe) alongside any published numbers.

\subsection{Leakage Audit Statement}
For the two prediction tasks the released sample definitions withhold the four ADAS-control variables (Section~\ref{sec:tasks}); the audit consists of the field-set disjointness check, the single-feature probe, and the label-reconstruction probe described in Appendix~\ref{app:protocol}.

\section{Baseline Configurations}
\label{app:baselines}

\subsection{Structured-Signal Baselines}
\textbf{LR / XGBoost.} Both operate on per-channel window statistics of the 50\,Hz structured signals (mean, standard deviation, minimum, maximum, first, last, and slope per channel). XGBoost uses 2{,}000 trees with early stopping (50 rounds on validation loss) and a validation-selected grid: max depth in $\{4,6,8\}$, learning rate in $\{0.03, 0.05\}$, minimum child weight in $\{1,5,20\}$, column subsampling in $\{0.6, 0.8\}$; class weighting is capped at 10, and the final model averages five reseeded fits. \textbf{GRU.} Separate modality branches with gated residual fusion. \textbf{TCN.} A temporal convolutional network on the same input.

\begin{table}[htbp]
\caption{Leakage-inflated full-input reference: single-seed model comparison and temporal ablation (cross-driver, $h{=}3$\,s) with the four ADAS-control variables included.}
\label{tab:model_comparison_appendix}
\centering
\small
\setlength{\tabcolsep}{3.2pt}
\renewcommand{\arraystretch}{1.02}
\begin{tabular*}{\columnwidth}{@{\extracolsep{\fill}}l cc cc cc@{}}
\toprule
\neurow & \multicolumn{2}{c}{Task 1} & \multicolumn{2}{c}{Task 2} & \multicolumn{2}{c}{Task 3} \\
\neurow Model & Acc & $F1_M$ & AUROC & AUPRC & AUROC & AUPRC \\
\midrule
LR & .865 & .838 & .812 & .609 & .783 & .350 \\
XGBoost (5\,s) & .936 & .920 & .828 & .631 & .877 & .653 \\
GRU (5\,s) & .926 & .910 & .815 & .590 & .843 & .429 \\
TCN & .925 & .911 & .770 & .554 & .838 & .472 \\
\midrule
XGBoost (last frame) & .790 & .700 & .782 & .449 & .870 & .608 \\
GRU (last frame) & .729 & .661 & .723 & .306 & .828 & .397 \\
\bottomrule
\end{tabular*}
\end{table}

\subsection{Fusion Models}
\textbf{Cross-Modal Transformer.} Per-modality token streams with cross-modal attention; $d{=}256$, 4 layers; the reported leak-safe runs use the structured stream.
\textbf{RG-HBT-Q} (7.4M parameters). A hierarchical fusion model that first encodes the controller signals grouped by semantic category, then fuses the three streams through a shared attention bottleneck following the attention-bottleneck design of \citet{nagrani2021mbt}, trained with modality dropout; in our robustness test, removing any single input stream costs it roughly half of what the same removal costs a plain Transformer.
\textbf{DI-RG-HBT-Q} (6.5--6.7M parameters). Extends RG-HBT-Q with per-sample driver-input event tokens that cross-attend the front-video, cabin-video, and remaining controller streams; it is the strongest purely neural multimodal model in our suite.

\subsection{World-Model Features}
Video features are extracted with frozen V-JEPA2 encoders~\cite{assran2025vjepa2} (ViT-L/256, 1024-d, and ViT-g/384, 1408-d) on 2\,s clips with a 0.5\,s stride for both cameras. On top of these latents we pretrain \textbf{LatentWM}, a 9.9M-parameter causal Transformer that predicts future clip latents at multiple horizons with heteroscedastic (mean and variance) heads; it is trained only on training-split routes under the leak-safe conditioning set. For each benchmark window we summarize the model's predictions, predictive uncertainty, and prediction error (``surprise'') into a fixed-length feature vector, which is consumed by the XGBoost readout. The ViT-g/384 variant is used where it robustly improved over ViT-L/256 (Task~2); Task~3 numbers use the ViT-L/256 pipeline.

\subsection{Driver-Pose Features}
A YOLO-based 17-keypoint pose estimator runs on the full in-cabin stream at 2\,fps. Each window is summarized by per-keypoint-channel mean, standard deviation, maximum, and last value. Coverage was audited after a full-route re-extraction: positive and negative windows have near-identical row coverage (about 10 rows per 5\,s window for both classes; zero-fraction $\le$1.5\% for both), ruling out a coverage artifact.

\subsection{\sysname-WM Composition}
\sysname-WM is a single XGBoost model per task with multimodal input features: for Task~2, window statistics $\oplus$ frozen world-model (ViT-g/384) video features; for Task~3, window statistics $\oplus$ driver-pose features. Both use the tuning grid and training protocol of Appendix~C.1; no model combination or probability mixing is involved.

\subsection{Long-Context and Distillation Models}
LCAT projects each stream (front and cabin V-JEPA2 tokens, 96-d pose, 45-d structured signals binned to 2\,Hz) to 256-d tokens with modality embeddings and rotary positional encoding, applies two causal self-attention layers per stream, fuses through a gated four-token attention bottleneck with four transition queries, and adds a wide tabular path over the long-window statistics (13.4--16.4M parameters, 15 epochs, validation-AUPRC selection). Distillation uses route-grouped five-fold out-of-fold teacher probabilities with weight 8; the privileged (LUPI) variant trains the teacher on leak-safe inputs including the override channels while the student input is checked to be strictly anticipation-safe.

\subsection{Hardware and Seeds}
All experiments run on a single NVIDIA RTX 5090. Multi-seed results use seeds $\{42, 123, 7\}$ (mean $\pm$ population standard deviation); the no-DMS specification in Table~\ref{tab:tenseed} uses ten seeds with no seed selection.

\section{Protocol Details}
\label{app:protocol}

\subsection{Driving Modes and Event Filters}
The automation state is active when \texttt{cc\_latActive} or \texttt{cruiseState\_enabled} equals one. A handover is an inactive-to-active transition and a takeover the reverse, after: a one-second state-persistence debounce; a minimum two-second automation episode and one-second human episode; merging of adjacent same-state episodes; and a minimum two-second gap between same-type events.

\subsection{Task-1 Action Rules and Validation}
Actions are labeled at 1\,Hz and aggregated by majority vote over each 5\,s window, with priority order Stopped $>$ Lane-change $>$ Turning $>$ Braking $>$ Accelerating $>$ Car-following $>$ Cruising (Table~\ref{tab:action_rules}). Perturbing each rule threshold by 20--30\% changes only 0.25--7.5\% of labels (Stopped 0.25--0.43\%, Turning 1.5--4.1\%, Accelerating 1.9--4.0\%, Braking 1.8--3.8\%, Car-following 2.2--7.5\%). A domain expert blindly annotated 500 class-balanced windows from video alone; agreement with the rule labels, computed only afterwards, is 96.0\% (Cohen's $\kappa=0.953$), with per-class F1 between 0.913 and 0.986 (lane-change 0.957). Because the same signal types that define the rules are also model inputs, part of the reported Task-1 performance is rule-aligned by construction; the rule-free ablation quantifies this directly: removing the nine rule-defining signals still yields a Macro-F1 of 0.577, and front-camera video alone (a modality the rules never use) reaches 0.592, both far above the seven-class chance level of 0.14 (Table~\ref{tab:rulefree}).

\begin{table}[htbp]
\caption{Task-1 action-labeling rules.}
\label{tab:action_rules}
\centering
\small
\setlength{\tabcolsep}{4pt}
\renewcommand{\arraystretch}{1.05}
\begin{tabular*}{\columnwidth}{@{\extracolsep{\fill}}p{0.22\columnwidth}p{0.70\columnwidth}@{}}
\toprule
\neurow Action & Rule \\
\midrule
Stopped & \texttt{vEgo} below 0.5\,m/s for at least 2\,s \\
Lane-change & \texttt{laneChangeState} $>$ 0, or a turn signal is on and steering angle exceeds 5$^\circ$ \\
Turning & steering angle above 10$^\circ$ for at least 1\,s \\
Braking & \texttt{aEgo} below $-0.35$\,m/s$^2$, or brake pressed \\
Accelerating & \texttt{aEgo} above 0.37\,m/s$^2$ for at least 1\,s \\
Car-following & lead vehicle active, gap below 60\,m, absolute acceleration below 1\,m/s$^2$ \\
Cruising & none of the above (default) \\
\bottomrule
\end{tabular*}
\end{table}

\subsection{Leak-Safe Field Audit}

\begin{table*}[htbp]
\caption{Per-category field audit for the leak-safe protocol.}
\label{tab:field_audit}
\centering
\small
\setlength{\tabcolsep}{5pt}
\renewcommand{\arraystretch}{1.05}
\begin{tabular*}{\textwidth}{@{\extracolsep{\fill}} p{0.30\textwidth} p{0.10\textwidth} p{0.12\textwidth} p{0.40\textwidth}}
\toprule
\neurow Signal category (examples) & Task 1 & Tasks 2/3 & Can it trivially reveal the target? \\
\midrule
Engagement flags (\texttt{cc\_latActive}, \texttt{cruiseState\_enabled}) & not used by rules & \textbf{removed} & Yes for Tasks 2/3: they define the label, hence removed. \\
Automation control state / command (\texttt{cs\_longControlState}, \texttt{actuators\_accel}) & not used by rules & \textbf{removed} & No (0\% within-window change; single-feature AUPRC at base rate); removed conservatively. \\
Automation warnings / take-over requests & not exposed & not exposed & N/A: no warning or request flag is exposed as input. \\
Planner outputs (\texttt{model\_desiredCurvature}, \texttt{model\_desiredAcceleration}) & not used by rules & kept & No (0\% within-window change; single-feature AUPRC at base rate). \\
Driver input (brake, accelerator, steering, blinkers) & partly rule signals (ablated) & kept & Not the label; strongest retained feature 0.247 vs.\ 0.332 for a removed flag. \\
Ego motion \& lane state (\texttt{vEgo}, \texttt{aEgo}, steering angle, lane state, lead distance) & define the action label (ablated) & kept & For Task 1, yes by construction; quantified (rule-free 0.577) and human-validated ($\kappa=0.953$). \\
\bottomrule
\end{tabular*}
\end{table*}

Table~\ref{tab:field_audit} lists every signal category and whether it can trivially reveal the transition label. For Tasks~2/3 only the four automation-control variables are removed; every retained signal is a genuine observation available at prediction time. For Task~1 no automation control, command, or warning signal participates in the action rules; the by-construction dependence on the rule-defining vehicle-state signals is quantified by the rule-free ablation (Table~\ref{tab:rulefree}) and validated by blind human annotation.

\subsection{Event-Level Metric}
Let an event's windows be all positive windows assigned to one transition. The event score is the maximum model score over those windows. Negatives are formed by splitting all remaining time into non-overlapping horizon-length bins, each scored by its maximum. Event-level AUPRC is computed over these event and bin scores. This removes the optimistic correlation between overlapping 0.5\,s-stride windows.

\subsection{Anticipation-Safe Protocol}
The anticipation-safe protocol (T3-A) applies two independent mechanisms on top of leak-safe. (i) \emph{Input withholding}: the six driver-override channels (\texttt{brakePressed}, \texttt{gasPressed}, \texttt{steeringPressed}, \texttt{steeringTorque}, \texttt{brake}, \texttt{gas}) are removed from the input in addition to the four ADAS-control variables. (ii) \emph{Window filtering}: for each takeover, the override onset is located as the first activation of a binary override signal within 3\,s before the event (threshold 0.5), and a positive window is kept only if it ends at least 1.0\,s \emph{before} that onset; system-initiated takeovers with no detectable override are cut at the event time. This filter reduces the $h{=}3$\,s test set from 217 to 92 positive events and the base rate from 0.120 to 0.035. Of the 1{,}787 benchmark takeovers with valid $h{=}3$\,s windows, 92.5\% have a detectable driver override within 3\,s (94.9\% in the cross-driver test split; the remainder are system-initiated). Together the two mechanisms guarantee that no part of the driver's takeover action is visible to the model. Results are in Table~\ref{tab:antsafe}; Table~\ref{tab:t3a_robust} reports buffer-sensitivity, and per-driver spreads and bootstrap CIs accompany the released analysis scripts.

\subsection{Negative Sampling}
For Tasks 2/3, positive windows slide at a 0.5\,s stride up to the event; negative window-ends slide at a 2.0\,s stride over transition-free state segments, excluding any $t$ with an event inside $[t-2, t+h+2]$ (a 2\,s buffer on both sides of the horizon). Negatives are capped per route at $\min(15 \cdot n_{\text{events}} \cdot h/0.5,\ 1000)$ and subsampled with a fixed seed (42). Windows within 5\,s of a route boundary are excluded. Hard negatives in which the driver acts but no state change follows are not specially curated; they occur naturally within the retained segments.

\subsection{VLM Evaluation Protocol}
We draw fixed evaluation sets of 350, 300, and 300 windows for Tasks 1--3 from the test split. Task~1 is class-balanced (50 windows per class), whereas Tasks 2 and 3 preserve the positive prevalence of the benchmark test split: the Task-2 subset contains 44 positives and 256 negatives (prevalence 0.147) and the Task-3 subset 36 positives and 264 negatives (0.120), so near-chance models score close to these prevalences in AUPRC. Each window provides three frames sampled at 0.5\,s, 2.5\,s, and 4.5\,s from each requested camera, and the Full configuration adds a structured text summary of vehicle state (speed, steering, pedals, lead distance) to front+cabin frames. Because this text overlaps the rule-defining signals of Task~1, T1-Full is a leakage-inflated diagnostic rather than an official rule-free Task-1 result; the official T1 comparison uses camera-only configurations. Models are prompted zero-shot to output a class label (Task~1) or a probability (Tasks~2/3); outputs are parsed with a fixed numeric parser, unparseable responses are excluded, and configurations with more than 15\% parse failures are marked with $*$. All models, ten open and two closed (GPT-4o and Gemini~2.5 Flash), are evaluated on the identical frozen benchmark-subset draw (seed 42) with the same prompts, frames, and parser, so every row is directly comparable. GPT-4o declines a large share of cabin-only and handover prompts (starred cells). The old ``All-modality'' GPS road-context text is not reproducible post-anonymization and is replaced by the Full configuration above.

\section{Complete Result Tables}
\label{app:vlm}

\subsection{Full VLM Sweep}

\begin{table*}[htbp]
\caption{Complete zero-shot VLM results (cross-driver, $h{=}3$\,s, sample-level, one frozen sample draw). $*$: over 15\% parse failures. T1 Full cells are leakage-inflated diagnostics.}
\label{tab:vlm_full}
\centering
\footnotesize
\setlength{\tabcolsep}{4pt}
\renewcommand{\arraystretch}{1.0}
\begin{tabular*}{\textwidth}{@{\extracolsep{\fill}} ll ccc | ll ccc}
\toprule
\neurow Model & Input & T1 $F1_M$ & T2 & T3 & Model & Input & T1 $F1_M$ & T2 & T3 \\
\midrule
Qwen2-VL-2B & Front & .131 & .147 & .131 & Qwen3-VL-4B-Th. & Front & .241 & .176$^*$ & .160 \\
Qwen2-VL-2B & Cabin & .084 & .147 & .146 & Qwen3-VL-4B-Th. & Cabin & .131$^*$ & .177$^*$ & .175$^*$ \\
Qwen2-VL-2B & F+C & .079 & .147 & .133 & Qwen3-VL-4B-Th. & F+C & .240 & .161$^*$ & .152 \\
Qwen2-VL-2B & Full & .058 & .147 & .151 & Qwen3-VL-4B-Th. & Full & .459 & .201$^*$ & .157$^*$ \\
\addlinespace[2pt]
Qwen2.5-VL-3B & Front & .083 & .147 & .122 & Qwen3-VL-8B & Front & .280 & .162 & .177 \\
Qwen2.5-VL-3B & Cabin & .058 & .149 & .138 & Qwen3-VL-8B & Cabin & .138 & .159 & .123 \\
Qwen2.5-VL-3B & F+C & .042 & .148 & .136 & Qwen3-VL-8B & F+C & .279 & .166 & .166 \\
Qwen2.5-VL-3B & Full & .056 & .147 & .123 & Qwen3-VL-8B & Full & .492 & .165 & .171 \\
\addlinespace[2pt]
Qwen2.5-VL-7B & Front & .227 & .161 & .120 & LLaVA-OV-1.5-8B & Front & .168 & .165 & .136 \\
Qwen2.5-VL-7B & Cabin & .109 & .141 & .126 & LLaVA-OV-1.5-8B & Cabin & .102 & .157 & .121 \\
Qwen2.5-VL-7B & F+C & .134 & .152 & .127 & LLaVA-OV-1.5-8B & F+C & .156 & .195 & .149 \\
Qwen2.5-VL-7B & Full & .383 & .147 & .118 & LLaVA-OV-1.5-8B & Full & .359 & .156 & .136 \\
\addlinespace[2pt]
Qwen3-VL-2B & Front & .194 & .187 & .121 & InternVL3.5-8B & Front & .142 & .154 & .146 \\
Qwen3-VL-2B & Cabin & .094 & .151 & .120 & InternVL3.5-8B & Cabin & .053 & .159 & .122 \\
Qwen3-VL-2B & F+C & .095 & .154 & .122 & InternVL3.5-8B & F+C & .123 & .154 & .161 \\
Qwen3-VL-2B & Full & .362 & .148 & .120 & InternVL3.5-8B & Full & .306 & .141 & .152 \\
\addlinespace[2pt]
Qwen3-VL-4B & Front & .209 & .166 & .165 & MiniCPM-V-4.5 & Front & .170 & .166 & .181 \\
Qwen3-VL-4B & Cabin & .069 & .155 & .132 & MiniCPM-V-4.5 & Cabin & .081 & .133 & .141 \\
Qwen3-VL-4B & F+C & .212 & .173 & .166 & MiniCPM-V-4.5 & F+C & .141 & .174 & .173 \\
Qwen3-VL-4B & Full & .456 & .173 & .167 & MiniCPM-V-4.5 & Full & .365 & .160 & .182 \\
\midrule
GPT-4o & Front & .272 & .141$^*$ & .162 & Gemini-2.5-Flash & Front & .362 & .184 & .173 \\
GPT-4o & Cabin & .381$^*$ & .147$^*$ & .120$^*$ & Gemini-2.5-Flash & Cabin & .163 & .163 & .154 \\
GPT-4o & F+C & .303 & .164 & .170 & Gemini-2.5-Flash & F+C & .360 & .170 & .184 \\
GPT-4o & Full & .586 & .165 & .196 & Gemini-2.5-Flash & Full & .619 & .183 & .202 \\
\bottomrule
\end{tabular*}
\end{table*}

Table~\ref{tab:vlm_full} reports all ten open models under all four modality configurations.

\subsection{Ten-Seed Comparison (no-DMS specification)}
Table~\ref{tab:tenseed} reports an independent robustness check at a single strict specification: leak-safe inputs with the in-cabin driver-monitoring group additionally removed, cross-driver split, $h{=}3$\,s, ten random seeds, no seed selection. It differs from the official protocol (which keeps the derived driver-monitoring outputs) and probes whether conclusions survive without them: the probability-level multimodal fusion beats the tabular baseline on handover and matches it on takeover.

\begin{table}[htbp]
\caption{Ten-seed leak-safe comparison without driver-monitoring inputs (sample-level AUPRC, mean $\pm$ std).}
\label{tab:tenseed}
\centering
\small
\setlength{\tabcolsep}{4pt}
\renewcommand{\arraystretch}{1.05}
\begin{tabular*}{\columnwidth}{@{\extracolsep{\fill}}l cc@{}}
\toprule
\neurow Model & Handover & Takeover \\
\midrule
XGBoost (controller signals) & 0.246 $\pm$ 0.007 & \textbf{0.441 $\pm$ 0.005} \\
V-JEPA2 video only & 0.241 $\pm$ 0.007 & 0.307 $\pm$ 0.010 \\
Multimodal fusion & \textbf{0.264 $\pm$ 0.008} & 0.440 $\pm$ 0.004 \\
\bottomrule
\end{tabular*}
\end{table}

\subsection{Generalization-Axis Results}

\begin{table}[htbp]
\caption{Cross-vehicle split (vehicle-model fingerprints held out; leak-safe, $h{=}3$\,s, sample\,/\,event AUPRC, 3-seed mean$\pm$std).}
\label{tab:xveh}
\centering
\small
\setlength{\tabcolsep}{4pt}
\renewcommand{\arraystretch}{1.05}
\begin{tabular*}{\columnwidth}{@{\extracolsep{\fill}}l cc@{}}
\toprule
\neurow Method & T2: Handover & T3-D: Takeover det. \\
\midrule
GRU & .337$\pm$.011 / .162$\pm$.007 & .436$\pm$.016 / .308$\pm$.014 \\
Cross-Modal Transformer & .302$\pm$.019 / .128$\pm$.013 & .333$\pm$.025 / .173$\pm$.018 \\
XGBoost (stats) & .360$\pm$.006 / .190$\pm$.008 & .443$\pm$.001 / .365$\pm$.003 \\
XGBoost (stats+pose) & .434$\pm$.006 / .257$\pm$.007 & .452$\pm$.002 / .362$\pm$.005 \\
\bottomrule
\end{tabular*}
\end{table}

\begin{table}[htbp]
\caption{Within-driver temporal split vs.\ the identity-holdout splits (leak-safe, $h{=}3$\,s, sample AUPRC, 3-seed mean$\pm$std).}
\label{tab:temporalsplit}
\centering
\small
\setlength{\tabcolsep}{4pt}
\renewcommand{\arraystretch}{1.05}
\begin{tabular*}{\columnwidth}{@{\extracolsep{\fill}} ll ccc}
\toprule
\neurow Task & Method & Temporal & Cross-driver & Cross-vehicle \\
\midrule
T2 & XGB stats & .454$\pm$.002 & .236$\pm$.003 & .360$\pm$.006 \\
T2 & XGB stats+pose & .465$\pm$.004 & .258$\pm$.003 & .434$\pm$.006 \\
T2 & GRU & .351$\pm$.020 & .202$\pm$.009 & .337$\pm$.011 \\
\midrule
T3-D & XGB stats & .532$\pm$.001 & .479$\pm$.007 & .443$\pm$.001 \\
T3-D & XGB stats+pose & .567$\pm$.001 & .514$\pm$.000 & .452$\pm$.002 \\
T3-D & GRU & .414$\pm$.028 & .280$\pm$.049 & .436$\pm$.016 \\
\bottomrule
\end{tabular*}
\end{table}

\subsection{World-Model and Pose Feature Ablation}
Table~\ref{tab:wm_ablation} decomposes the \sysname-WM gains. World-model features drive the Task-2 gain (0.236$\rightarrow$0.335 with the ViT-g/384 encoder), while pose features drive the Task-3 gain (0.479$\rightarrow$0.514); combining all three feature blocks in one tabular model does not help further.

\begin{table}[htbp]
\caption{Feature-block ablation for the XGBoost readout (leak-safe, cross-driver, $h{=}3$\,s; sample\,/\,event AUPRC; 3-seed mean $\pm$ std, single seed where marked).}
\label{tab:wm_ablation}
\centering
\footnotesize
\setlength{\tabcolsep}{3pt}
\renewcommand{\arraystretch}{1.05}
\begin{tabular*}{\columnwidth}{@{\extracolsep{\fill}}l cc@{}}
\toprule
\neurow Features & Task 2 & Task 3 \\
\midrule
stats & .236$\pm$.003 / .103 & .479$\pm$.007 / .380 \\
stats + pose & .258$\pm$.003 / .114 & .514$\pm$.000 / .413 \\
stats + WM (ViT-L/256) & .306$\pm$.005 / .148 & .452$\pm$.001 / .352 \\
stats + WM (ViT-g/384) & .335$\pm$.003 / .171 & .460$\pm$.002 / .360 \\
WM only (ViT-L/256) & .265$\pm$.008 / .130 & .280$\pm$.006 / .166 \\
WM only (ViT-g/384) & .286$\pm$.000 / .165 & .365$\pm$.002 / .242 \\
stats + WM + pose & .303$\pm$.008 / .140 & .477$\pm$.003 / .373 \\
Distilled DI-RG-HBT-Q & .266$\pm$.005 / .098 & .468$\pm$.011 / .334 \\
\bottomrule
\end{tabular*}
\end{table}

\subsection{Anticipation-Safe Results}
Table~\ref{tab:antsafe} reports Task~3 under the anticipation-safe protocol. All feature sets stay near the base rate, i.e., current performance on T3-D largely reflects early \emph{detection} of the driver's override rather than anticipation before it starts.

\begin{table}[htbp]
\caption{T3-A robustness: buffer sensitivity (3-seed) and event-level bootstrap 95\% CIs (seed 42, 2000 resamples).}
\label{tab:t3a_robust}
\centering
\small
\setlength{\tabcolsep}{4pt}
\renewcommand{\arraystretch}{1.05}
\begin{tabular*}{\columnwidth}{@{\extracolsep{\fill}} l ccc}
\toprule
\neurow Pre-override buffer & Sample AUPRC & Event AUPRC & Events \\
\midrule
0.5\,s & .067$\pm$.002 & .031$\pm$.001 & 96 \\
1.0\,s (official) & .056$\pm$.001 & .026$\pm$.001 & 92 \\
1.5\,s & .043$\pm$.001 & .023$\pm$.001 & 87 \\
2.0\,s & .030$\pm$.001 & .019$\pm$.000 & 76 \\
\midrule
\neurow Bootstrap 95\% CI & stats & stats+pose & \\
\midrule
T3-D event AUPRC & .379 [.315, .441] & .411 [.343, .474] & \\
T3-A event AUPRC & .026 [.019, .041] & .036 [.025, .065] & \\
\bottomrule
\end{tabular*}
\end{table}

\subsection{Second-Set Consistency and Evaluation-Set Growth}
The driver-disjoint validation split contains 15{,}326 anticipation-safe windows with 590 positives across 160 events (base rate 0.039), a set larger than the test split's 92 events. The same models behave identically there: XGBoost statistics reaches 0.075$\pm$.004 sample / 0.036$\pm$.004 event AUPRC and stats+pose 0.102$\pm$.003 / 0.058$\pm$.002, against 0.522 / 0.365 for detection on the same split. Because this split was used for hyperparameter selection, any bias favors higher anticipation scores, which strengthens the near-chance reading. Anticipation events accrue linearly with driving time at roughly 5--6 test events per test-split hour, and the pre-override filter retains roughly 40--42\% of takeovers, so enlarging the anticipation evaluation is a data-collection matter rather than a protocol limitation: reaching about 300 test events, which would tighten the bootstrap CIs by roughly 1.8$\times$, requires growing the corpus to about 500--550 hours. The corpus grew from 129 to 204.9 hours in the three months since its initial release, and expanding the anticipation evaluation set is the primary data target for the next benchmark version.

\subsection{Anticipation Attempts Beyond the Standard Baselines}
Table~\ref{tab:t3a_attempts} lists every additional attempt at T3-A. All context lengths, architectures, and training signals stay near the base rate, while the same tools work on T3-D, which is why we attribute the gap to missing information rather than to modeling choices.

\begin{table}[htbp]
\caption{Additional T3-A attempts (cross-driver, $h{=}3$\,s, sample AUPRC, 3-seed mean$\pm$std; base rate 0.035; best standard baseline 0.070).}
\label{tab:t3a_attempts}
\centering
\small
\setlength{\tabcolsep}{4pt}
\renewcommand{\arraystretch}{1.05}
\begin{tabular*}{\columnwidth}{@{\extracolsep{\fill}} l c}
\toprule
\neurow Attempt & T3-A AUPRC \\
\midrule
Long-window stats (XGB), 30\,s & .064$\pm$.002 \\
Long-window stats (XGB), 60\,s & .068$\pm$.002 \\
LCAT, video streams only, 60\,s & .039$\pm$.003 \\
LCAT, pose stream only, 60\,s & .033$\pm$.002 \\
LCAT, all streams, 30\,s & .047$\pm$.003 \\
LCAT, all streams, 60\,s & .061$\pm$.004 \\
LCAT + teacher distillation (LUPI, privileged teacher) & .070$\pm$.004 \\
\midrule
Same LCAT + distillation on T3-D (reference) & .367$\pm$.014 \\
Long-window statistics on T3-D, 60\,s (reference) & .497 \\
\bottomrule
\end{tabular*}
\end{table}

\subsection{Deployed-Predictor Baseline Details}

Disengagement predictions were parsed from the original openpilot logs (modelV2 messages at 20\,Hz) for the 26 benchmark routes whose raw logs remain available, with alignment verified against the CAN stream (median speed correlation 1.000). Scores use the 4\,s-horizon combined disengagement probability, maximized over each window. On the 5 covered test routes: T3-D 1{,}458 windows / 178 positives / 30 events, deployed predictor 0.293\,/\,0.101 vs.\ XGBoost statistics 0.771\,/\,0.732 on identical rows; T3-A has 9 positives and 3 events, too few for a stable estimate, and no method separates from chance there.

\subsection{Transition-Subtype Stratification}
Classifying each test event by which assistance flag changed: on T2, the model recalls lateral-only activations far better than longitudinal-only ones (0.90 vs.\ 0.66 at the operating threshold; per-subtype event AUPRC 0.20 vs.\ 0.02). On T3-D the concentration is stronger: lateral-only takeovers reach 0.72 event AUPRC (recall 0.79) while simultaneous lateral+longitudinal takeovers, the most safety-relevant full disengagements, reach only 0.05 (recall 0.30). Pooled metrics therefore overstate performance on full takeovers, a clear target for future methods.

\begin{table}[htbp]
\caption{Task-3 takeover under the anticipation-safe protocol (driver-override channels withheld; sample\,/\,event AUPRC; base rate 0.035).}
\label{tab:antsafe}
\centering
\small
\setlength{\tabcolsep}{4pt}
\renewcommand{\arraystretch}{1.05}
\begin{tabular*}{\columnwidth}{@{\extracolsep{\fill}} l c}
\toprule
\neurow Features (XGBoost readout) & AUPRC \\
\midrule
stats & .056$\pm$.001 / .026 \\
stats + pose & .070$\pm$.001 / .040 \\
stats + WM (ViT-L/256) & .043$\pm$.001 / .019 \\
stats + WM (ViT-g/384) & .049$\pm$.001 / .024 \\
\bottomrule
\end{tabular*}
\end{table}

\subsection{Task-1 Rule-Free Ablation}
Table~\ref{tab:rulefree} reports the Task-1 input ablation supporting the signal-derived framing in Section~\ref{sec:tasks}.

\begin{table}[htbp]
\caption{Task-1 ablation (GRU, cross-driver): performance without the rule-defining signals and from modalities the rules never use.}
\label{tab:rulefree}
\centering
\small
\setlength{\tabcolsep}{4pt}
\renewcommand{\arraystretch}{1.05}
\begin{tabular*}{\columnwidth}{@{\extracolsep{\fill}} l ccc}
\toprule
\neurow Input & $F1_M$ & LC F1 & Acc \\
\midrule
Full structured (with rule signals) & 0.889 & 0.833 & 0.903 \\
Rule-free structured & 0.577 & 0.344 & 0.632 \\
Front-camera video only & 0.592 & 0.106 & 0.693 \\
IMU only & 0.352 & 0.101 & 0.391 \\
Driver-monitoring only & 0.219 & 0.016 & 0.271 \\
Chance (seven classes) & 0.14 & -- & -- \\
\bottomrule
\end{tabular*}
\end{table}

\subsection{Task-1 Multi-Label Variant}
Removing the priority order and evaluating each rule independently per second (window label = per-class majority) yields the released multi-label variant with the same 1.16M windows. Per-class positive rates: Car-following 40.3\%, Braking 21.8\%, Stopped 19.0\%, Turning 16.6\%, Accelerating 11.3\%, Lane-change 1.3\%; 31.2\% of windows carry two or more labels (e.g., Stopped+Car-following 4.5\%, Braking+Car-following 4.0\%, Turning+Braking 2.2\%), and 30.6\% carry none (cruising). This quantifies how much behavioral co-occurrence the single-label priority ordering discards. Table~\ref{tab:task1ml} reports GRU baselines (binary cross-entropy, 3 seeds) on rule-free inputs: both modalities recover the frequent classes well, and front video, which the rules never use, matches the structured input on turning and acceleration while lane-change remains hard from video alone.

\begin{table}[htbp]
\caption{Task-1 multi-label baselines (GRU, cross-driver, rule-free inputs, per-class and macro average precision; 3-seed mean$\pm$std).}
\label{tab:task1ml}
\centering
\small
\setlength{\tabcolsep}{4pt}
\renewcommand{\arraystretch}{1.05}
\begin{tabular*}{\columnwidth}{@{\extracolsep{\fill}}l cc@{}}
\toprule
\neurow Class (label rate) & Structured (rule-free) & Front video \\
\midrule
Stopped (19.0\%) & .919$\pm$.011 & .908$\pm$.004 \\
Lane-change (1.3\%) & .461$\pm$.055 & .065$\pm$.013 \\
Turning (16.6\%) & .718$\pm$.027 & .736$\pm$.038 \\
Braking (21.8\%) & .758$\pm$.021 & .593$\pm$.027 \\
Accelerating (11.3\%) & .599$\pm$.018 & .732$\pm$.012 \\
Car-following (40.3\%) & .978$\pm$.004 & .908$\pm$.002 \\
\midrule
Macro-AP & .739$\pm$.006 & .657$\pm$.015 \\
Micro-F1 / Macro-F1 @0.5 & .750 / .638 & .748 / .629 \\
\bottomrule
\end{tabular*}
\end{table}

\subsection{Operating-Point Evaluation (False Alarms and Lead Time)}

\begin{table*}[htbp]
\caption{Operating-point evaluation (cross-driver, $h{=}3$\,s, seed 42). R@$f$: event recall at $f$ false alarms per driving hour; lead columns are at 1\,FA/h.}
\label{tab:leadtime}
\centering
\small
\setlength{\tabcolsep}{5pt}
\renewcommand{\arraystretch}{1.05}
\begin{tabular*}{\textwidth}{@{\extracolsep{\fill}} ll ccc ccc c}
\toprule
\neurow Protocol & Model & R@0.5 & R@1 & R@2 & R@1, lead$\ge$1s & lead$\ge$2s & lead$\ge$3s & Median lead (s) \\
\midrule
T2 & XGB stats & .032 & .037 & .046 & .032 & .018 & .005 & 2.0 \\
\rowcolor{neuGray!35}
T2 & XGB stats+WM(g384) & .041 & .073 & .115 & .069 & .041 & .005 & 2.0 \\
T3-D & XGB stats & .267 & .281 & .295 & .277 & .240 & .124 & 3.0 \\
\rowcolor{neuGray!35}
T3-D & XGB stats+pose & .267 & .277 & .323 & .272 & .217 & .115 & 3.0 \\
T3-A & XGB stats & .000 & .000 & .000 & .000 & .000 & .000 & -- \\
\rowcolor{neuGray!35}
T3-A & XGB stats+pose & .011 & .022 & .022 & .022 & .011 & .000 & 2.25 \\
\bottomrule
\end{tabular*}
\end{table*}

Table~\ref{tab:leadtime} evaluates warning behavior at fixed false-alarm budgets. The decision threshold is swept so that the number of alarmed negative horizon-bins per monitored driving hour ($\approx$17.4\,h of test-set driving for T2, 16.1\,h for T3) equals the budget; an event counts as recalled if any of its windows scores above the threshold, and as recalled with lead $\ge L$ if such a window ends at least $L$ seconds before the transition. The median lead is over the earliest alarmed window per recalled event and is capped by the 3\,s horizon. Even the best models catch a small fraction of events at practical false-alarm rates, which we consider the most deployment-relevant summary of the current state of the benchmark.

\clearpage

\end{document}